\PassOptionsToPackage{unicode}{hyperref}
\PassOptionsToPackage{hyphens}{url}
\PassOptionsToPackage{dvipsnames,svgnames,x11names}{xcolor}
\documentclass[
  12pt]{article}

\usepackage{amsmath,amssymb,amsthm}
\usepackage{iftex}
\ifPDFTeX
  \usepackage[T1]{fontenc}
  \usepackage[utf8]{inputenc}
  \usepackage{textcomp} 
\else 
  \usepackage{unicode-math}
  \defaultfontfeatures{Scale=MatchLowercase}
  \defaultfontfeatures[\rmfamily]{Ligatures=TeX,Scale=1}
\fi
\usepackage{lmodern}
\ifPDFTeX\else
\fi
\IfFileExists{upquote.sty}{\usepackage{upquote}}{}
\IfFileExists{microtype.sty}{
  \usepackage[]{microtype}
  \UseMicrotypeSet[protrusion]{basicmath} 
}{}
\makeatletter
\@ifundefined{KOMAClassName}{
  \IfFileExists{parskip.sty}{%
    \usepackage{parskip}
  }{
    \setlength{\parindent}{0pt}
    \setlength{\parskip}{6pt plus 2pt minus 1pt}}
}{
  \KOMAoptions{parskip=half}}
\makeatother
\usepackage{xcolor}
\makeatletter
\ifx\paragraph\undefined\else
  \let\oldparagraph\paragraph
  \renewcommand{\paragraph}{
    \@ifstar
      \xxxParagraphStar
      \xxxParagraphNoStar
  }
  \newcommand{\xxxParagraphStar}[1]{\oldparagraph*{#1}\mbox{}}
  \newcommand{\xxxParagraphNoStar}[1]{\oldparagraph{#1}\mbox{}}
\fi
\ifx\subparagraph\undefined\else
  \let\oldsubparagraph\subparagraph
  \renewcommand{\subparagraph}{
    \@ifstar
      \xxxSubParagraphStar
      \xxxSubParagraphNoStar
  }
  \newcommand{\xxxSubParagraphStar}[1]{\oldsubparagraph*{#1}\mbox{}}
  \newcommand{\xxxSubParagraphNoStar}[1]{\oldsubparagraph{#1}\mbox{}}
\fi
\makeatother

\usepackage{longtable,booktabs,array}
\usepackage{calc} 
\usepackage{etoolbox}
\makeatletter
\patchcmd\longtable{\par}{\@ifnoskipsec\mbox{}\fi\par}{}{}
\makeatother
\IfFileExists{footnotehyper.sty}{\usepackage{footnotehyper}}{\usepackage{footnote}}
\makesavenoteenv{longtable}
\usepackage{graphicx}
\makeatletter
\def\maxwidth{\ifdim\Gin@nat@width>\linewidth\linewidth\else\Gin@nat@width\fi}
\def\maxheight{\ifdim\Gin@nat@height>\textheight\textheight\else\Gin@nat@height\fi}
\makeatother
\setkeys{Gin}{width=\maxwidth,height=\maxheight,keepaspectratio}
\makeatletter
\def\fps@figure{htbp}
\makeatother

\makeatletter
\@ifpackageloaded{caption}{}{\usepackage{caption}}
\AtBeginDocument{%
\ifdefined\contentsname
  \renewcommand*\contentsname{Table of contents}
\else
  \newcommand\contentsname{Table of contents}
\fi
\ifdefined\listfigurename
  \renewcommand*\listfigurename{List of Figures}
\else
  \newcommand\listfigurename{List of Figures}
\fi
\ifdefined\listtablename
  \renewcommand*\listtablename{List of Tables}
\else
  \newcommand\listtablename{List of Tables}
\fi
\ifdefined\figurename
  \renewcommand*\figurename{Figure}
\else
  \newcommand\figurename{Figure}
\fi
\ifdefined\tablename
  \renewcommand*\tablename{Table}
\else
  \newcommand\tablename{Table}
\fi
}
\@ifpackageloaded{float}{}{\usepackage{float}}
\floatstyle{ruled}
\@ifundefined{c@chapter}{\newfloat{codelisting}{h}{lop}}{\newfloat{codelisting}{h}{lop}[chapter]}
\floatname{codelisting}{Listing}

\makeatother
\makeatletter
\@ifpackageloaded{caption}{}{\usepackage{caption}}
\@ifpackageloaded{subcaption}{}{\usepackage{subcaption}}
\makeatother

\ifLuaTeX
  \usepackage{selnolig}  
\fi
\usepackage[]{natbib}
\usepackage{bookmark}

\IfFileExists{xurl.sty}{\usepackage{xurl}}{} 
\newcommand{\papertitle}{Space-filling foldover designs for order-of-addition experiments under Kendall tau distance criteria}
\hypersetup{
  pdftitle={\papertitle},
  pdfauthor={Anonymous},
  pdfkeywords={Permutation search; Pairwise ordering; Mallows kernel; simulated annealing; Bayesian optimization; surrogate modeling},
  colorlinks=true,
  linkcolor={blue},
  filecolor={Maroon},
  citecolor={Blue},
  urlcolor={Blue},
  pdfcreator={LaTeX via pandoc}}

\usepackage{bm}
\usepackage{multirow}
\usepackage[english]{babel}
\usepackage{tipa}
\usepackage{lscape}
\usepackage{epsfig}
\usepackage{epstopdf}

\usepackage{algorithm}
\usepackage{algpseudocode}
\usepackage{setspace}
\AtBeginEnvironment{algorithmic}{\setstretch{1.05}}
\algrenewcommand\algorithmicindent{1.5em}
\algrenewcommand\textproc{\textsf}
\algrenewcommand\algorithmicrequire{\textbf{Input:}}
\algrenewcommand\algorithmicensure{\textbf{Output:}}
\algrenewcommand\algorithmiccomment[1]{\hfill\(\triangleright\) #1}
\makeatletter
\renewcommand{\ALG@name}{Algorithm}
\def\ALG@capt@plain{\relax}
\makeatother

\theoremstyle{plain}
\newtheorem{theorem}{Theorem}

\newtheorem{corollary}{Corollary}
\newtheorem{proposition}{Proposition}

\theoremstyle{definition}
\newtheorem{definition}{Definition}
\newtheorem{example}{Example}

\theoremstyle{remark}
\newtheorem{remark}{Remark}
\newtheorem*{remark*}{Remark}

\newcommand{\ba}{\begin{array}}
\newcommand{\ea}{\end{array}}
\newcommand{\bt}{\begin{tabular}}
\newcommand{\et}{\end{tabular}}
\newcommand{\btb}{\begin{table}}
\newcommand{\etb}{\end{table}}
\newcommand{\bc}{\begin{center}}
\newcommand{\ec}{\end{center}}
\newcommand{\bea}{\begin{eqnarray}}
\newcommand{\eea}{\end{eqnarray}}
\newcommand{\Bea}{\begin{eqnarray*}}
\newcommand{\Eea}{\end{eqnarray*}}
\newcommand{\beq}{\begin{equation}}
\newcommand{\eeq}{\end{equation}}

\def\spacingset#1{\renewcommand{\baselinestretch}%
{#1}\small\normalsize}

\def \bfm#1{\mbox{\boldmath$#1$}}

\def \x {{\bfm x}} 
 \def \y {{\bfm y}}

\newcommand{\ave}{{\rm ave}}

\newcommand{\anon}{1}
\newcommand{\paperauthorblock}{%
Hui Shao, Yaping Wang, Dongdong Xiang, \\
KLATASDS--MOE, School of Statistics, \\
East China Normal University, Shanghai 200062, China \\
and \\
Qian Xiao \\
Department of Statistics, School of Mathematical Sciences, \\
Shanghai Jiao Tong University, Shanghai 200240, China \\
}

\begin{document}

\def\spacingset#1{\renewcommand{\baselinestretch}%
{#1}\small\normalsize} \spacingset{1}


\if1\anon
{
 \title{\bf \papertitle}
  \author{\paperauthorblock}
  \maketitle
} \fi

\if0\anon
{
  \bigskip
  \bigskip
  \bigskip
  \begin{center}
    {\LARGE\bf \papertitle}
\end{center}
  \medskip
} \fi

\bigskip
\begin{abstract}
Order-of-addition experiments are used when the response depends on the sequence in which components are introduced.
The full design consists of all permutations and is therefore infeasible for even moderately sized component sets.
We study fractional order-of-addition designs from a space-filling point of view, using the Kendall tau distance to measure discrepancies in pairwise precedence relations.
Our criteria include the minimum pairwise Kendall tau distance, the average distance, and the second moment of the pairwise distance distribution.
We connect these criteria to the pairwise ordering model and to Gaussian process prediction with the Mallows kernel.
To efficiently construct space-filling designs, we develop a Foldover Simulated Annealing algorithm based on the Kendall tau Distance (FSA-KD). This algorithm searches over a representative half-design, exploits incremental distance updates, and optimizes a well-justified weighted criterion.
Numerical studies demonstrate that the resulting designs have favorable space-filling properties, 
improve prediction accuracy under pairwise ordering and Mallows-kernel Gaussian process models, and perform well in permutation-space Bayesian optimization tasks.
\end{abstract}

\noindent%

\vfill

\newpage
\spacingset{1.8} 

\section{Introduction}\label{sec1}

Order-of-addition (OofA) experiments are used when the response of a system depends on the sequence in which a set of components is introduced.
Such problems arise in many applications, including drug combination studies, chemical engineering, job scheduling, and other permutation-based optimization problems \citep{wang2020,voelkel2019,zhao2021}; see \citet{linpeng2019} and \citet{linrios2025} for recent reviews.
For an OofA experiment with $m$ components,  the design space consists of all \(m!\) permutations.
This factorial growth makes the full design infeasible even for moderate $m$.
For example, when \(m=10\), the full design contains more than \(3.6\) million runs.
Thus, a central task in OofA experimentation is to select a small set of informative orders that can support modeling, prediction, or the search for a desirable order.

Several statistical models and design criteria have been proposed for OofA experiments.
The pairwise ordering (PWO) model represents an order through the
relative precedence of component pairs and has become one of the most commonly used models \citep{vannostrand1995,peng2019,mee2020,wanglin2023}.
Position-based models, including the component-position (CP) and flexible-position models, instead describe orders through the positions occupied by the components \citep{yang2021,stokes2022}.
Gaussian process and other surrogate models have also been considered when the response surface over the permutation space is highly nonlinear or when sequential optimization is desired  \citep{jiao2015,XiaoXu21,deshwal2022}.
These developments have motivated model-based criteria such as \(D\)-, \(A\)-, \(E\)-, and MS-optimality, as well as prediction-oriented criteria  \citep{stokes2024metaheuristic,XiaoWangMandalDeng24}.

Good fractional OofA designs can be obtained through either algebraic constructions or algorithmic search.
Two representative algebraic structures are order-of-addition orthogonal arrays (OofA-OAs) and component orthogonal arrays (COAs), which provide balance and optimality properties under PWO and position-based models \citep{peng2019,yang2021,zhao2021,schoenmee2023}.
 More recently, \citet{tsai2025} characterized OofA-OAs through centralized generalized wordlength patterns, providing a useful minimum aberration viewpoint for comparing fractional OofA designs.
 \citet{tsai2025} recently characterized OofA-OAs through centralized generalized
wordlength patterns.
These constructions, however, may restrict the allowable run size, number of components, or algebraic structure.
 Algorithmic methods provide a more flexible alternative.
Existing searches, including exchange algorithms, threshold accepting, particle swarm optimization, and differential evolution, are typically guided by model-based criteria or position-based distances such as the Hamming, \(L_1\), and \(L_2\) distances \citep{voelkel2019,li2022,stokes2024,stokes2024metaheuristic,huang2025}.
These distances are useful for position-based models, but they do not directly measure discrepancies in pairwise precedence,
which motivates a distance criterion more closely aligned with the PWO representation.

These considerations motivate the use and study of Kendall tau distance, which counts the number of component pairs whose relative orders differ between two permutations \citep{kendall1938} and is therefore directly aligned with PWO coding.
The aim of this article is to construct and study space-filling fractional OofA designs through three summaries of the pairwise Kendall tau distances: the minimum, the average, and the second moment.
These criteria describe worst-case separation, overall separation, and dispersion of the distance distribution, respectively.
They also have direct statistical interpretations.
We show that, under the PWO model, the average distance and second moment determine the MS-optimality criterion and correspond to the first two entries of the centralized generalized wordlength pattern of \citet{tsai2025}.
In addition, under a Gaussian process model with the Mallows kernel \citep{jiao2015,deshwal2022}, the maximin criterion is further supported by a maximum-entropy argument when correlation decays rapidly with distance.

For efficient construction of space-filling OofA designs, we propose a foldover simulated annealing algorithm based on the Kendall tau distance, denoted by FSA-KD \citep{metropolis1953,kirkpatrick1983,aarts1988}.
For even run sizes, FSA-KD searches over a representative half-design of size \(h=n/2\) and obtains the remaining runs by reversal.
This strategy reduces the search dimension and makes the Kendall tau criterion amenable to fast incremental updates.
Foldover identities reduce the cost of evaluating a proposed move from \(O(n^2m^2)\) under full recomputation to \(O(hm^2)\), while local swaps reduce it further to \(O(hm)\) by changing only \(O(m)\) pairwise-ordering indicators.
We show that the foldover structure also fixes the average Kendall tau distance, so the search focuses on a well-defined normalized weighted criterion involving the minimum distance and the second moment.
For odd run sizes, a simple deletion step provides a practical extension.

We evaluate FSA-KD in both design construction and downstream statistical tasks.
In our implementation, the algorithm constructs designs with up to \(m=20\) components and \(n=100\) runs in less than one second.
Across the settings considered, FSA-KD produces larger minimum Kendall tau distances and higher values of the proposed composite criterion than random sampling and metaheuristic methods based on position distances.
The resulting designs also perform favorably in sparse-PWO model prediction,
Mallows-kernel Gaussian process prediction, and Bayesian optimization over permutation spaces.
These results indicate that the computational savings do not come at the expense of design quality.

The remainder of this paper is organized as follows.
Section~\ref{sec2} introduces the Kendall tau distance-based space-filling criteria.
Section~\ref{sec3} gives their statistical and design-theoretic justifications under the PWO model, the centralized wordlength pattern, and the Gaussian process model with the Mallows kernel.
Section~\ref{sec4} develops the foldover construction and the proposed FSA-KD algorithm.
Numerical studies and comparisons are given in Section~\ref{sec5}, and Section~\ref{sec6} concludes the paper with some discussions.
Technical proofs and additional numerical results are provided in the Supplementary Materials.

\section{Kendall tau distance-based space-filling criteria}\label{sec2}

Let $\mathcal{Z}_m=\{0,1,\ldots,m-1\}$ be the set of components, and let
$\mathcal{S}_m$ denote the set of all permutations of $\mathcal{Z}_m$.
An $n$-run order-of-addition design for $m$ components is represented by an
$n\times m$ array
$
D=\{\x_1,\ldots,\x_n\}=(x_{ir})_{n\times m},
$
where each row $\x_i=(x_{i1},\ldots,x_{im})\in\mathcal{S}_m$ gives one addition
order. Here, the index $r\in\{1,\ldots,m\}$ denotes the position in the sequence, and
$x_{ir}$ is the component placed in position $r$ of run $i$. For a component
$c\in\mathcal{Z}_m$, define the position map $\pi_{\x_i}(c)$ by
\[
\pi_{\x_i}(c)=r \quad \Longleftrightarrow \quad x_{ir}=c.
\]
Thus $\pi_{\x_i}(c)$ is the position of component $c$ in the order $\x_i$.
The full OofA design is $D_{\mathrm{full}}=\mathcal{S}_m$ and has $m!$ runs.
Since $m!$ increases rapidly with $m$, full designs are rarely feasible in practice except
for small component sizes. This motivates the construction of small-sized fractional OofA
designs with desirable properties.

Designs for OofA experiments may be evaluated either by model-based criteria or by model-free criteria \citep{linrios2025}.
Model-based criteria are tied to a specified regression model and usually optimize a function of the information matrix, such as the determinant in the D-optimality criterion. Model-free criteria, such as space-filling criteria, instead assess how well the selected runs are spread over the design space.
They are therefore useful when the working model is uncertain or when the design is intended for surrogate modeling. The maximin distance criterion \citep{johnson1990} is a standard space-filling criterion.
For OofA designs, distances based on component positions, such as the Hamming, $L_1$, and $L_2$ distances, have been used in the literature \citep{li2022,stokes2024metaheuristic,huang2025}.
In this paper, we focus on the Kendall tau distance \citep{kendall1938}, which measures discrepancies in pairwise precedence relations and is therefore well matched to OofA experimental runs.

For two runs $\x_i$ and $\x_j\in\mathcal{S}_m$, their Kendall tau distance is defined as
\begin{equation}\label{eq:kendall-position}
  k(\x_i,\x_j)
  = \sum_{0\le a<b\le m-1}
  \mathbf{1}{\left\{\bigl(\pi_{\x_i}(a)-\pi_{\x_i}(b)\bigr)
  \bigl(\pi_{\x_j}(a)-\pi_{\x_j}(b)\bigr)<0\right\}},
\end{equation}
where $ \mathbf{1}{\{\cdot \}}$ is the indicator function.
By \eqref{eq:kendall-position}, $k(\x_i,\x_j)$ is the number of component pairs whose relative order is
reversed between $\x_i$ and $\x_j$. It follows that
$
0\le k(\x_i,\x_j)\le \binom{m}{2},
$
and $k(\x_i,\x_j) =0$ if and only if $\x_i=\x_j$, and $k(\x_i,\x_j) = \binom{m}{2}$
if and only if the two orders are reverses of one another, that is,
$x_{jr}=x_{i,m+1-r}$ for $r=1,\ldots,m$.

The space-filling quality of an OofA design can be summarized through the
multiset of its inter-run Kendall tau distances,
$
\{k(\x_i,\x_j):1\le i<j\le n\}.
$
We consider three space-filling criteria based on $
\{k(\x_i,\x_j):1\le i<j\le n\}$. The first is the minimum pairwise
Kendall tau distance,
\begin{equation}\label{kmin}
k_{\min}(D)=\min_{1\le i<j\le n} k(\x_i,\x_j).
\end{equation}
This is the Kendall tau analogue of the classical maximin $L_p$-distance criterion.
An OofA design is called a maximin Kendall tau distance design if it maximizes
$k_{\min}(D)$ among all $n$-run OofA designs with $m$ components.

The second criterion is the average Kendall tau distance,
\begin{equation}\label{kave}
k_{\ave}(D)=\frac{1}{\binom{n}{2}}\sum_{1\le i<j\le n}k(\x_i,\x_j).
\end{equation}
A larger value of $k_{\ave}(D)$ indicates greater average separation among the
selected orders.
For traditional U-type designs (i.e., orthogonal array of strength one),  the corresponding average
$L_p$ distance is a constant; see, for example, \citet{ZhouXu15}. This property, however,
does not hold for OofA designs under the Kendall tau distance.
Even a balanced OofA design, in the sense that each component appears equally
often in each position, need not have a fixed average Kendall tau distance; see Example~\ref{ex1}.
This is one reason for treating $k_{\ave}(D)$ as a separate design criterion.

The third metric is the second moment of the pairwise Kendall tau distances,
\begin{equation}\label{km2}
  k_{m_2}(D)
  = \frac{1}{\binom{n}{2}} \sum_{1\le i<j\le n} k^2(\x_i,\x_j).
\end{equation}
Second-moment criteria have been used to study distance distributions of
space-filling designs; see, for example, \citet{wang2022}
for U-type designs under $L_p$ distances. For OofA designs, $k_{m_2}(D)$ should be interpreted
together with $k_{\ave}(D)$, because the average Kendall tau distance is not
fixed in general. In particular, when two designs have the same value of
$k_{\ave}(D)$, minimizing $k_{m_2}(D)$ is equivalent to minimizing the dispersion
of the pairwise distances, since
\[
\sum_{1\le i<j\le n}\{k(\x_i,\x_j)-k_{\ave}(D)\}^2
=
\sum_{1\le i<j\le n}k^2(\x_i,\x_j)
-\binom{n}{2}k_{\ave}^2(D).
\]
Thus $k_{\min}(D)$ controls the worst-case separation, $k_{\ave}(D)$ measures
global separation, and $k_{m_2}(D)$ describes the concentration of the distance
distribution once the average separation is accounted for.

\begin{example}\label{ex1}
Consider the following two $4\times 4$ OofA designs:
\[
D_1=
\begin{pmatrix}
0&1&2&3\\
1&2&3&0\\
2&3&0&1\\
3&0&1&2
\end{pmatrix},
\qquad
D_2=
\begin{pmatrix}
0&1&2&3\\
1&3&0&2\\
2&0&3&1\\
3&2&1&0
\end{pmatrix}.
\]
Both designs are balanced in the component-position sense. However, their
six pairwise Kendall tau distances are
\[
\begin{array}{c|cccccc}
 \textrm{Row pairs} & (1,2)&(1,3)&(1,4)&(2,3)&(2,4)&(3,4)\\
\hline
D_1 & 3&4&3&3&4&3\\
D_2 & 3&3&6&6&3&3
\end{array}.
\]
Thus the two designs have the same minimum distance,
$
k_{\min}(D_1)=k_{\min}(D_2)=3,
$
but different average distances:
$
k_{\ave}(D_1)={10}/{3}$ and $k_{\ave}(D_2)=4$.
They also have different second moments,
$
k_{m_2}(D_1)={34}/{3},
k_{m_2}(D_2)=18.
$
Under the $k_{\ave}$ criterion, the design $D_2$ is better, although it has a larger $k_{m_2}$ value than $D_1$.
\end{example}

\section{Statistical and design-theoretic justifications}\label{sec3}

The Kendall tau distance measures introduced in Section~\ref{sec2} are
model-free geometric criteria.  This section explains why these geometric
summaries also matter statistically.  The first argument uses the pairwise
ordering model, and the second uses a Gaussian process model with a Mallows
kernel.  We also relate the distance criteria to the centralized generalized
wordlength pattern proposed by \citet{tsai2025}, which clarifies how they
compare with existing OofA balance conditions.

\subsection{Pairwise ordering model}

A commonly used model for OofA experiments is the PWO model \citep{vannostrand1995,voelkel2019,peng2019}.
For a run $\x\in\mathcal{S}_m$ and a component pair $0\le a<b\le m-1$, define
\[
z_{ab}(\x)=
\begin{cases}
+1, & \pi_\x(a)<\pi_\x(b),\\
-1, & \pi_\x(a)>\pi_\x(b).
\end{cases}
\]
That is, $z_{ab}(\x)$ records whether component $a$ precedes component $b$ in run $\x$.
Let
\begin{equation}\label{eq:pwo-vector}
z(\x)=(z_{ab}(\x):0\le a<b\le m-1)^\top
\end{equation}
be the $q$-dimensional pairwise ordering vector, where
$q=\binom{m}{2}$.  The PWO model is
\begin{equation}\label{eq:pwo}
y(\x)=\beta_0+\sum_{0\le a<b\le m-1}\beta_{ab}z_{ab}(\x)+\varepsilon,
\end{equation}
where the errors are independent with mean zero and variance $\sigma^2$.

For an $n$-run design $D=\{\x_1,\ldots,\x_n\}$, let
\[
X=
\begin{pmatrix}
1 & z(\x_1)^\top\\
\vdots & \vdots\\
1 & z(\x_n)^\top
\end{pmatrix}
\]
be the model matrix under \eqref{eq:pwo}, and let $M=X^\top X$ be the moment matrix.  Under normal errors, this matrix is proportional to the Fisher information matrix.  The
MS-optimality criterion of \citet{eccleston1974} is to minimize
$\operatorname{tr}(M^2)$, where $\operatorname{tr}(\cdot)$ is the trace function; see also \cite{peng2019}.
The next result shows that this criterion  is determined by $k_{\ave}(D)$ and $k_{m_2}(D)$ defined in \eqref{kave} and \eqref{km2}, respectively.

\begin{theorem}\label{th:pwoms}
Let $D = \{\x_1,\ldots,\x_n\}$ be an OofA design for studying $m$ components.  Under the
PWO model \eqref{eq:pwo}, the MS-optimality criterion function
\begin{equation}\label{eq:th:pwoms}
\operatorname{tr}(M^2)
=
4n(n-1)k_{m_2}(D)
-\{2m(m-1)+4\}n(n-1)k_{\ave}(D)
+\frac{n^2(m^2-m+2)^2}{4}.
\end{equation}
\end{theorem}

Theorem~\ref{th:pwoms} gives a direct link between the proposed Kendall tau distance
metrics ($k_{m_2}$ and $k_{\ave}$) and a model-based criterion.  For fixed $n$ and $m$, a larger
average Kendall tau distance and a smaller second moment tend to reduce
$\operatorname{tr}(M^2)$.  Thus the MS-optimality criterion favors designs whose pairwise
ordering vectors are well separated on average but whose distance distribution
is not overly dispersed.

The same identity also gives a useful lower bound.  Let $M_{\textrm{full}}$ be the information
 matrix of the full OofA design under the PWO model, i.e., $M_{\textrm{full}} =   X_{\textrm{full}}^\top X_{\textrm{full}} $, where $X_{\textrm{full}}$ is the PWO model matrix of the full design.
By Theorem~2 of
\citet{peng2019}, the eigenvalues of the normalized (per-run) information matrix $ (m!)^{-1} M_{\textrm{full}} $ are
$
1, {(m+1)}/{3}$, and ${1}/{3}$,
with multiplicities $1$, $m-1$, and $\binom{m-1}{2}$, respectively.  Consequently,
by the optimality of the full-design moment matrix \citep[Theorem~1]{peng2019}, we have
\[
\frac{\operatorname{tr}(M^2)}{n^2}
\ge
\frac{\operatorname{tr}(M_{\textrm{full}}^2)}{(m!)^2} = \frac{1}{18}\{2m^3+3m^2-5m+18\},
\]
and equality holds if and only if the normalized information matrix $M/n$ coincides with that  of the full OofA design.
Combining this inequality with Theorem~\ref{th:pwoms} yields the following result.

\begin{corollary}\label{coro2}
For any $n$-run ($n\ge 2$) OofA design $D$ with $m$ components, we have
$$k_{m_2}(D)
\ge
\frac{m^2-m+2}{2}\,k_{\ave}(D)
-\frac{nm}{144(n-1)}
\left(9m^3-22m^2+39m-26\right).
$$
The equality holds if and only if the design $D$ has the same normalized information matrix as the full design.
\end{corollary}

Corollary~\ref{coro2} shows that the two distance moments cannot be optimized independently: for fixed $n$ and $m$, a larger average Kendall tau distance raises the lower bound on $k_{m_2}(D)$, with equality attained at the full-design information benchmark. Thus, $k_{m_2}(D)$ should be interpreted together with $k_{\ave}(D)$, rather than as a stand-alone smaller-the-better measure.

\subsection{Connection with centralized generalized wordlength patterns}
\label{subsec:cgwlp}

The preceding subsection shows that the first two moments of the
Kendall tau distance distribution are directly linked to the MS criterion
under the PWO model. A related viewpoint is provided by the centralized
generalized wordlength pattern of \citet{tsai2025}, which compares a
fractional OofA design with the full design through normalized
\(J\)-characteristics of the PWO model matrix.

Let
$
\mathcal{Q}=\{(a,b):0\le a<b\le m-1\}$,  $q=|\mathcal{Q}|=\binom{m}{2}.
$
For a subset $W\subseteq\mathcal{Q}$, define
$
z_W(\x)=\prod_{(a,b)\in W}z_{ab}(\x),
$
with $z_\emptyset(\x)=1$.  The corresponding $J$-characteristic of an OofA design
$D$ is
\[
J_W(D)=\sum_{x\in D}z_W(\x).
\]
Let $D_{\mathrm{full}}$ denote the full design with $m!$ runs.  Following
\citet{tsai2025}, the centralized generalized
wordlength pattern is defined by
\begin{equation}\label{eq:C-def}
C_a(D)=
\sum_{|W|=a}
\left\{
\frac{J_W(D)}{n}-\frac{J_W(D_{\mathrm{full}})}{m!}
\right\}^2,
  a=1,\ldots,q.
\end{equation}
The quantity $C_a(D)$ measures  how close the normalized \(J\)-characteristics of
\(D\) are to those of the full design when projected onto subsets of \(a\) pairwise
ordering factors.

An $n\times m$ OofA design $D$ is called an order-of-addition orthogonal array of strength $t$, denoted by OofA-OA$(n,m,t)$ if, for every
$W\subseteq\mathcal{Q}$ with $|W|\le t$, the frequency distribution of the
submatrix $\{z_{ab}(\x):\x\in D,\ (a,b)\in W\}$ is proportional to that of the
corresponding submatrix of the full design $D_{\mathrm{full}}$ \citep{zhao2021,schoenmee2023}.
By \citet{tsai2025}, $D$ is an OofA-OA$(n,m,t)$ if and only if
$C_a(D)=0$ for $a=1,\ldots,t$.
Equivalently, an OofA-OA matches the full permutation design up to strength \(t\)
in terms of normalized \(J\)-characteristics.
For a general design $D$, the generalized minimum aberration criterion is to sequentially minimize
$C_a(D)$ for $a=1,\ldots,q$ \citep{xuwu2001}.

The next proposition gives a direct relationship between the first two
centralized wordlength quantities and the Kendall tau distance moments.

\begin{proposition}\label{prop:C-distance}
For any $n$-run OofA design $D=\{\x_1,\ldots,\x_n\}$ with $m$ components,
\begin{equation*}
C_1(D)
=
q-\frac{2(n-1)}{n}k_{\ave}(D),
\end{equation*}
and
\begin{equation*}\label{eq:C2-km2}
C_2(D)
=
\frac{q(q-1)}{2}
-\frac{2q(n-1)}{n}k_{\ave}(D)
+\frac{2(n-1)}{n}k_{m_2}(D)
-\frac{m(m-1)(m-2)}{18}.
\end{equation*}
\end{proposition}

Proposition~\ref{prop:C-distance} shows that the first two centralized
wordlength quantities are equivalent to the first two moments of the
Kendall tau distance distribution. The quantity $C_1(D)$ determines the
average Kendall tau distance, while $C_2(D)$ determines the second moment once
$C_1(D)$ is fixed. Thus the distance-based criteria in Section~\ref{sec2} and
the centralized wordlength viewpoint describe the same low-order structure of
the PWO representation.

Proposition~\ref{prop:C-distance} also gives a direct link to the MS-optimality criterion in
Theorem~\ref{th:pwoms}.

\begin{corollary}\label{coro3}
For any $n$-run OofA design $D$ with $m$ components, we have
$$\operatorname{tr}(M^2)
=
n^2\left\{
 (m!)^{-2} \operatorname{tr}(M_{\textrm{full}}^2)+2C_1(D)+2C_2(D)
\right\}.
$$
\end{corollary}
By Corollary~\ref{coro3}, the MS-optimality criterion over the full-design benchmark is exactly
determined by the first two entries of the centralized wordlength pattern.

As an immediate consequence of Proposition~\ref{prop:C-distance}, if $D$ is an OofA-OA$(n,m,2)$, then
$C_1(D)=C_2(D)=0$, and hence
\begin{equation*}
k_{\ave}(D)
=
\frac{nm(m-1)}{4(n-1)}
\end{equation*}
and
\begin{equation*}
k_{m_2}(D)
=
\frac{nm(m-1)(9m^2-5m+10)}{144(n-1)}.
\end{equation*}
These identities explain why strength-two OofA-OAs attain the full-design
benchmark for the first two Kendall tau distance moments.

Another class of designs is the component orthogonal array (COA) \citep{yang2021}.  An $n\times m$
OofA design $D$ is called a COA, denoted by COA$(n,m)$,
if for any two columns  every ordered pair $(a,b)$ with $a\ne b$ appears
equally often.  This pair-position balance implies that  $C_1(D)=0$, and by
Proposition~\ref{prop:C-distance},
\[
k_{\ave}(D)=\frac{nq}{2(n-1)}
=
\frac{nm(m-1)}{4(n-1)}.
\]
Thus COA$(n,m)$ designs attain the same full-design benchmark as
OofA-OA$(n,m,2)$ designs for the average Kendall tau distance.  However, the
COA condition does not generally determine $C_2(D)$, and therefore does not
fix $k_{m_2}(D)$.

Exact OofA-OAs of strength $t\ge 2$ and COAs are available only for restricted
combinations of $n$ and $m$, and their run sizes are often large when $m$
increases (at least of order $O(m^2)$).  This motivates the more flexible foldover construction developed
in Section~\ref{sec4}.

\subsection{Gaussian process model with the Mallows kernel}

The Kendall tau distance also arises naturally in kernel models for
permutation-valued inputs.  \citet{jiao2015} introduced the Mallows kernel as
a positive definite kernel on permutations based on the Kendall tau distance,
and \citet{deshwal2022} used related kernels in Gaussian process (GP) models for
optimization over permutation spaces.  Since an OofA run is itself a
permutation of the components, the Mallows kernel provides a natural way to
model the correlation between two addition orders through their Kendall tau
distance.
In this subsection, we show that the maximin Kendall tau criterion is also supported by a  D-optimality argument under a GP surrogate model.

Consider the GP model
\begin{equation}\label{eq:mallows_gp}
Y(\x)=\mu+Z(\x),
\end{equation}
where $\mu$ is an unknown constant and $Z(\x)$ is a zero-mean GP
with covariance function
$
\operatorname{cov}\{Z(\x_i),Z(\x_j)\}
=
\sigma^2 K(\x_i,\x_j; \theta).
$
We use the Mallows kernel
\begin{equation}\label{eq:mallows_kernel}
K(\x_i,\x_j;\theta)
=\exp\{-\theta \cdot k(\x_i,\x_j)\},
\end{equation}
where \(k(\x_i,\x_j)\) is the Kendall tau distance between \(\x_i\) and \(\x_j\).
The parameter $\theta > 0$ controls the rate at which the correlation decreases
as the distance increases.

For an $n$-run design $D=\{\x_1,\ldots,\x_n\}$, let
$\Sigma_D(\theta)$ be the $n\times n$ kernel correlation matrix with entries
$K(\x_i,\x_j;\theta), 1\le i,j\le n.$
The D-optimality criterion considered here is to maximize
$\det\{\Sigma_D(\theta)\}$, which favors designs whose runs are weakly
correlated under the chosen kernel.  We use this determinant as a
maximum-entropy criterion for the GP covariance structure; it is not a
regression-information criterion for estimating the constant mean $\mu$.
This criterion is closely related to maximum-entropy designs for GP models \citep{silvey1980,santner2018}.
The following result shows that, when the kernel correlation decays rapidly with
the Kendall tau distance, the determinant criterion is governed by the closest
pairs in the design. This provides a permutation-space analogue of the
maximin $L_p$-distance argument of \citet{johnson1990}.

For $r=0,\ldots,q$, where $q=\binom{m}{2}$, define
$
\nu_r(D)=\#\{(i,j):1\le i<j\le n,\ k(\x_i,\x_j)=r\}.
$

\begin{theorem}\label{thm:maximinGP-Dopt}
Fix $m$ and $n$, and consider $n$-run OofA designs with no replicated runs.
Let $D^*$ be a maximin Kendall tau distance design, and write
$
k^*=k_{\min}(D^*).
$
Suppose that, among all designs attaining the maximum value $k^*$,
$D^*$ has the smallest number of pairs at distance $k^*$, that is,
$\nu_{k^*}(D^*)$ is minimal among all maximin Kendall tau distance designs.
Then, under the Gaussian process model \eqref{eq:mallows_gp} with kernel
\eqref{eq:mallows_kernel}, $D^*$ is   asymptotically
D-optimal as $\theta\to\infty$.  More precisely, for any competing design
$D$ satisfying either
$
k_{\min}(D)<k^*
$
or
$
k_{\min}(D)=k^*
\text{ and }
\nu_{k^*}(D)>\nu_{k^*}(D^*),
$
we have
$$
\det\{\Sigma_{D^*}(\theta)\}
>
\det\{\Sigma_D(\theta)\}
$$
for all sufficiently large $\theta$.
\end{theorem}

\begin{remark}\label{rem:tie}
The strict inequality in Theorem~\ref{thm:maximinGP-Dopt} should not be
claimed for every competing design.  If another design has the same value of
$k_{\min}$ and the same number of pairs attaining this distance, the leading
terms in the determinant expansion are identical.  In that case, the
first-order argument cannot rank the two designs.  A further comparison would
require the next distance level or higher-order terms in the determinant
expansion.
\end{remark}

 The regime $\theta\to\infty$ corresponds to a rapidly decaying kernel where only  permutations that are very close in Kendall tau distance remain strongly correlated, and the determinant of the kernel matrix is mainly
affected by the closest pairs in the design.  Theorem~\ref{thm:maximinGP-Dopt}
therefore gives a GP-model-based justification for the maximin Kendall tau distance
criterion.  For moderate values of $\theta$, however, the determinant also
depends on the remaining distance levels, so the theorem should be read as an
asymptotic motivation rather than a finite-$\theta$ characterization.
Together with the PWO results in the previous subsections, this shows that the
Kendall tau distance is not only a natural geometric measure on permutation
spaces, but also has direct links to  commonly used model-based statistical criteria.

\section{Construction of space-filling foldover OofA designs}\label{sec4}

In this section, we develop a foldover-based construction method for
space-filling OofA designs under the Kendall tau distance. Since the
permutation design space has size $m!$, exhaustive search over all addition
orders becomes computationally prohibitive even for moderate $m$. To reduce
the search burden while retaining useful distance properties, we introduce a
foldover structure for OofA designs and establish its main properties under
the Kendall tau metric. Building on these properties, we propose a foldover
simulated annealing algorithm for Kendall tau distance designs, referred to as
the {\em FSA-KD} algorithm.

\subsection{Distance properties of foldover designs}
\label{sec:foldover_distance}

Foldover constructions are classical tools in experimental design
\citep{hedayat1999}.  In the present OofA setting, the foldover of an addition
order is defined as its reverse order.  This transformation changes the sign of every PWO
factor and gives a simple symmetry in the Kendall tau distance.

For even $n$, write $n=2h$ and let
\[
D=H\cup \widetilde H,
\qquad
H=\{\x_1,\ldots,\x_h\},
\qquad
\widetilde H=\{\widetilde{\x}_1,\ldots,\widetilde{\x}_h\}.
\]
Here $H$ is the representative half-design and $\widetilde H$ is its foldover
counterpart.

\begin{definition}[Foldover transformation]
\label{def:foldover}
For a permutation $\x=(x_1,\ldots,x_m)\in\mathcal{S}_m$, its foldover
permutation $\widetilde{\x}$ is defined by
\begin{equation*}
\widetilde{x}_r=x_{m+1-r},\qquad r=1,\ldots,m.
\end{equation*}
\end{definition}

By Definition~\ref{def:foldover}, the foldover transformation reverses the relative order of
every pair of components. Hence
\(
z(\widetilde{\x})=-z(\x),
\)
where $z(\x)$ is the PWO vector defined in \eqref{eq:pwo-vector}. Conversely,
if a permutation $\y\in\mathcal{S}_m$ satisfies $z(\y)=-z(\x)$, then every
pairwise order in $\x$ is reversed in $\y$, which forces
$\y=(x_m,\ldots,x_1)$. Thus the reverse-order mapping is the unique
permutation whose PWO vector is $-z(\x)$.

Let $q=\binom{m}{2}$ be the maximum possible Kendall tau distance between two
orders.  The next theorem summarizes the distance properties of foldover OofA
designs.

\begin{theorem}
\label{th:foldover}
Let $D=H\cup\widetilde H$ be an $n=2h$-run foldover OofA design for $m$
components, where $H=\{\x_1,\ldots,\x_h\}$ and $q=\binom{m}{2}$.
Suppose that $h\ge 2$ and that $D$ has no repeated run; equivalently, $H$
contains no duplicated run and no pair of foldover runs.  Then the following
results hold.

\smallskip
\noindent\textup{(i)} For $1\le i<j\le h$,
\[
k(\widetilde{\x}_i,\widetilde{\x}_j)=k(\x_i,\x_j),
\qquad
k(\x_i,\widetilde{\x}_i)=q,
\]
and, for $i\ne j$,
\[
k(\x_i,\widetilde{\x}_j)=q-k(\x_i,\x_j).
\]
Thus all pairwise Kendall tau distances in $D$ are determined by the distance
matrix of $H$.

\smallskip
\noindent\textup{(ii)} The average Kendall tau distance is
\[
k_{\ave}(D)
=
\frac{nm(m-1)}{4(n-1)}.
\]

\smallskip
\noindent\textup{(iii)}  The minimum Kendall tau distance satisfies
\[
1\le k_{\min}(D)
\le
\left\lfloor\frac{m(m-1)}{4}\right\rfloor.
\]

\smallskip
\noindent\textup{(iv)} The second moment satisfies
\[
L_2\le k_{m_2}(D) \le U_2,
\]
where
\begin{equation}
\label{eq:C2_def}
L_2=
\frac{nm(m-1)(9m^2-5m+10)}{144(n-1)},
\end{equation}
and
\begin{equation}
\label{eq:U2_def}
U_2=
\frac{n m^2(m-1)^2-4(n-2)\{m(m-1)-2\}}{8(n-1)},
\end{equation}
The lower bound $L_2$ is attained
by OofA-OA$(n,m,2)$ designs whenever such designs exist.
\end{theorem}

\begin{remark}
\label{rem:C2U2}
The lower bound \(L_2\) defined by \eqref{eq:C2_def} is exact for designs with the normalized full PWO information matrix, but it may not be attainable within designs with small run sizes. The upper bound $U_2$ defined in \eqref{eq:U2_def} is
generally conservative because equality would require all representative-pair
distances to be either \(1\) or \(q-1\), which is usually impossible when \(h\)
is moderate or large. Nevertheless, these bounds provide normalization
benchmarks for defining the multi-objective space-filling criteria in the next
subsection.
\end{remark}

Theorem~\ref{th:foldover} shows why the foldover class is useful for
constructing space-filling OofA designs.
Only $h=n/2$ representative permutations need to be searched, and the average Kendall tau distance is fixed at the benchmark attained by the full design, OofA-OA$(n,m,2)$ designs, and COA$(n,m)$ designs.
The theorem also gives explicit bounds for the minimum distance and the second distance moment.  These properties provide useful benchmarks for evaluating candidate OofA designs and motivate the search procedure developed below.

\subsection{FSA-KD algorithm}

Based on the foldover properties in Theorem~\ref{th:foldover}, we propose a foldover simulated annealing algorithm based on the Kendall tau distance, denoted by FSA-KD, for constructing space-filling OofA designs.  For even run sizes $n=2h$, FSA-KD searches over the
representative half-design
$
H=\{\x_1,\ldots,\x_h\},
$
and then completes the design by the foldover operation,
$
D=H\cup\widetilde H.
$
The search is restricted to half-designs for which $H\cup\widetilde H$ has no
repeated run.

We first define the objective function used in the FSA-KD algorithm.
By Theorem~\ref{th:foldover}\textup{(ii)}, $k_{\ave}(D)$ is fixed within the class of nonreplicated foldover designs.
Therefore, the two remaining distance criteria to be optimized are \(k_{\min}(D)\) and \(k_{m_2}(D)\).
The former is to be maximized, whereas the latter is to be minimized.
Moreover, the two quantities have different numerical scales.
To make their weighted combination
meaningful, we put both criteria on a  $0$-to-$1$ desirability
scale.

Let
$
B_1=\left\lfloor {m(m-1)}/{4}\right\rfloor$,
and let $L_2$ and $U_2$ be defined in \eqref{eq:C2_def} and
\eqref{eq:U2_def}, respectively.  For any nonreplicated foldover design $D$,
Theorem~\ref{th:foldover} shows
\[
1\le k_{\min}(D)\le B_1,
\qquad
L_2\le k_{m_2}(D)\le U_2.
\]
Thus
$
{(k_{\min}(D)-1)}/{(B_1-1)}
$
measures the relative improvement of the minimum distance over its smallest
possible value, whereas
$
{(U_2-k_{m_2}(D))}/{(U_2-L_2)}
$
measures the relative improvement of the second moment over its upper
benchmark.  Both scaled quantities are larger for more desirable designs.  We
therefore define the weighted criterion
\begin{equation}
\label{eq:compoundobj}
\Phi_\lambda(D)
=
\lambda
\frac{k_{\min}(D)-1}{B_1-1}
+
(1-\lambda)
\frac{U_2-k_{m_2}(D)}{U_2-L_2},
\qquad 0\le \lambda \le 1.
\end{equation}
The weight $\lambda$ controls the relative emphasis on the maximin and second-moment components.  The
case $\lambda=1$ gives the GP-motivated maximin Kendall tau criterion in
Theorem~\ref{thm:maximinGP-Dopt}.  The case $\lambda=0$ gives the
second-moment criterion.  Since $k_{\ave}(D)$ is fixed within the foldover
class, this endpoint is also equivalent to minimizing the MS criterion in
Theorem~\ref{th:pwoms}.  Intermediate values of $\lambda$ give compromise
designs.  We use $\lambda=0.5$ as the default choice in the numerical studies,
which gives equal weight to the two normalized objectives.  Other values may
be used when one criterion is considered more important.

When $m=3$, $B_1=1$ and the normalized minimum-distance term is degenerate.
In that special case, the first term in \eqref{eq:compoundobj} may be omitted;
in the nontrivial cases considered below, $m\ge4$ and $B_1>1$.

FSA-KD uses two types of neighborhood moves.  The first is a global
replacement move, in which one representative order $\x_r$ is replaced by a
 randomly generated permutation.  The second is a local swap move, in which
two positions within $\x_r$ are exchanged.  In both cases, the foldover
counterpart is updated deterministically.  The two moves serve different
purposes.  The global replacement move helps the algorithm explore different regions of the permutation space, whereas the local move refines the current half-design through small changes.
The move type is chosen according to the current temperature $T$.
Specifically, the probability of a global replacement move is $T/T_0$, and the
probability of a local swap move is $1-T/T_0$.  Thus the search is more exploratory at high temperatures and becomes more local as the temperature decreases.

The foldover structure gives a simple and efficient update of the distance criteria.
A direct recomputation of all pairwise Kendall tau distances in an unrestricted \(n\)-run design requires $O(n^2m^2)$ operations per iteration when the distance is evaluated from all component pairs.
 In FSA-KD, only one order in the half-design is modified at each iteration.  Therefore,
only its distances to the remaining $h-1$ representative orders need to be
updated.  The distances involving foldover counterparts are then obtained from
\[
k(\x_i,\widetilde{\x}_j)=q-k(\x_i,\x_j),
\qquad q=\binom{m}{2}, \quad i\ne j,
\]
together with \(k(\x_i,\widetilde{\x}_i)=q\).
The per-iteration distance update cost is therefore reduced to
$O(hm^2)=O(nm^2)$.
As an indication of the practical cost, under the 500-evaluation budget used in Section~\ref{sec5-2}, a single-core implementation on a MacBook Air with an Apple M3 chip takes approximately 0.03, 0.11, and 0.79 seconds for \((m,n)=(5,25),(10,50)\), and \((20,100)\), respectively.

For implementation, it is enough to store the distance matrix within  the representative
half-design $H$.  By Theorem~\ref{th:foldover},
\[
k_{\min}(D)
=
\min_{1\le i<j\le h}
\min\{k(\x_i,\x_j),\,q-k(\x_i,\x_j)\},
\]
and
\[
k_{m_2}(D)
=
\frac{
h q^2+
2\sum_{1\le i<j\le h}
\left[
k^2(\x_i,\x_j)+\{q-k(\x_i,\x_j)\}^2
\right]
}
{h(2h-1)}.
\]
Suppose that a candidate half-design $H'$ is obtained from $H$ by replacing
only $\x_r$ with $\x'_r$.  All distances not involving $\x_r$ remain unchanged.
Let
$
g(u)=u^2+(q-u)^2.
$
Then the change in the second-moment numerator is
\[
\Delta G_r
=
\sum_{j\ne r}
\left[
g\{k(\x'_r,\x_j)\}-g\{k(\x_r,\x_j)\}
\right],
\]
and hence
\[
k_{m_2}(D')
=
k_{m_2}(D)
+
\frac{2\Delta G_r}{h(2h-1)}.
\]
Similarly,
\[
\begin{aligned}
k_{\min}(D')
=
\min\Bigg\{&
\min_{\substack{1\le i<j\le h\\ i,j\ne r}}
\min\{k(\x_i,\x_j),q-k(\x_i,\x_j)\},\\
&
\min_{j\ne r}
\min\{k(\x'_r,\x_j),q-k(\x'_r,\x_j)\}
\Bigg\}.
\end{aligned}
\]
In practice, this update is implemented by removing the old values
\(
\min\{k(\x_r,\x_j),q-k(\x_r,\x_j)\}, j\ne r,
\)
and inserting the new values
\(
\min\{k(\x'_r,\x_j),q-k(\x'_r,\x_j)\}, j\ne r,
\)
before taking the minimum.

For a global replacement move, the distances $k(\x'_r,\x_j)$, $j\ne r$, are recomputed directly.  For a local swap move, they can be updated more cheaply.
Suppose $\x'_r$ is obtained from $\x_r$ by swapping the entries in
positions $s<t$.  Only the following component pairs change their relative orders:
\[
\mathcal P_{st}
=
\{\{x_{rs},x_{rt}\}\}
\cup
\{\{x_{rs},x_{r\ell}\},\{x_{rt},x_{r\ell}\}:s<\ell<t\}.
\]
For an unordered pair $\{u,v\}$, write
$
z_{\{u,v\}}(\x)=z_{\min(u,v),\max(u,v)}(\x).
$
Then, for each $j\ne r$,
\[
k(\x'_r,\x_j)
=
k(\x_r,\x_j)
+
\sum_{\{u,v\}\in\mathcal P_{st}}
z_{\{u,v\}}(\x_r)z_{\{u,v\}}(\x_j).
\]
Indeed, for each pair in $\mathcal P_{st}$, the PWO sign in $\x_r$ is
reversed.  The Kendall tau distance to $\x_j$ increases by one if the pair
originally agrees with $\x_j$, and decreases by one if it originally
disagrees with $\x_j$.  Hence a local swap updates only
$2(t-s)-1$ pairwise ordering relations, rather than all
$q=\binom{m}{2}$ relations.

Combining the above updates gives the objective increment
\[
\Delta\Phi_\lambda
=
\lambda
\frac{k_{\min}(D')-k_{\min}(D)}{B_1-1}
-
(1-\lambda)
\frac{2\Delta G_r}{h(2h-1)(U_2-L_2)}.
\]
The candidate design is accepted if
$\Delta\Phi_\lambda>0$; otherwise, it is accepted with probability
\(
\exp\{\Delta\Phi_\lambda/T\}.
\)
This simulated annealing rule allows occasional downhill moves and helps avoid
premature convergence to a local optimum.
The proposed FSA-KD algorithm is summarized in Algorithm~\ref{alg:SAKD}.
Unless otherwise stated, the numerical studies use $T_0=1.0$, $T_{\min}=10^{-8}$,
$\alpha=0.997$, $N_{\max}=6000$, and $\lambda=0.5$.
An implementation-oriented version with additional details is provided in the section S3 of Supplementary Materials.

\begin{algorithm}[ht!]
\caption{FSA-KD algorithm for space-filling foldover OofA designs}
\label{alg:SAKD}
\begin{algorithmic}[1]
\Require Number of components $m$, even run size $n=2h$, weight $\lambda$, and annealing schedule $(T_0,\alpha,T_{\min},N_{\max})$
\Ensure An optimized foldover OofA design $D^*$
\State Initialize $H=\{\x_1,\ldots,\x_h\}$ such that $H\cup\widetilde H$ has no repeated run
\State Set $H^*\leftarrow H$, $\Phi^*\leftarrow\Phi_\lambda(H\cup\widetilde H)$, and $T\leftarrow T_0$
\For{$t=1,\ldots,N_{\max}$ while $T\ge T_{\min}$}
    \State Generate a candidate $H'$ by changing one representative order in $H$
    \If{$H'\cup\widetilde H'$ has no repeated run}
        \State Compute $\Delta\Phi_\lambda=\Phi_\lambda(H'\cup\widetilde H')-\Phi_\lambda(H\cup\widetilde H)$
        \If{$\Delta\Phi_\lambda>0$ or $\operatorname{rand}()<\exp\{\Delta\Phi_\lambda/T\}$}
            \State Set $H\leftarrow H'$
            \State If $\Phi_\lambda(H\cup\widetilde H)>\Phi^*$, set $H^*\leftarrow H$ and update $\Phi^*$
        \EndIf
    \EndIf
    \State Set $T\leftarrow \alpha T$
\EndFor
\State Return $D^*=H^*\cup\widetilde H^*$
\end{algorithmic}
\smallskip \noindent\footnotesize\textit{Note.} Here $\operatorname{rand}()$ denotes an independent uniform draw from $[0,1]$. \end{algorithm}

Although FSA-KD is designed for foldover designs and therefore naturally gives even run sizes, it can be adapted to odd run sizes by a deletion step.
For an odd target run size $n$, we first construct an
$(n+1)$-run foldover design using Algorithm~\ref{alg:SAKD}.
We then delete each row in turn and retain the $n$-run design with the largest \(k_{\min}\), breaking ties by the smaller \(k_{m_2}\).
The resulting design is no longer exactly foldover, but it inherits the separation structure of the parent foldover design and provides a practical extension to odd run sizes.

\section{Numerical results}\label{sec5}

This section evaluates the proposed FSA-KD algorithm from both design and
modeling perspectives.  We first examine the weight parameter \(\lambda\) in
the composite criterion \(\Phi_\lambda(D)\) defined in
\eqref{eq:compoundobj}.
We then compare FSA-KD with the simple random sampling (SRS) method and with the particle swarm optimization (PSO) and differential evolution (DE) algorithms of \citet{stokes2024metaheuristic}.
The latter two methods optimize a position-based maximin-\(L_2\) criterion and serve as efficient metaheuristic baselines.
Finally, we examine whether the resulting designs are useful for the
model-based tasks that motivated Section~\ref{sec3}, including prediction
under the PWO and Mallows-kernel GP models and Bayesian optimization over
permutation spaces.

Unless otherwise stated, FSA-KD uses the default schedule and weight specified in Section~\ref{sec4}.
For each design-size setting, all stochastic methods
are run independently for the stated number of repetitions, and the reported values are averages over these repetitions.  The comparisons use a common evaluation budget whenever possible, because the competing algorithms have different internal update rules.
The design criteria are the minimum pairwise distance \(k_{\min}(D)\), the average distance \(k_{\ave}(D)\), and the second moment \(k_{m_2}(D)\).
We also report the normalized components
\[
A(D)=\frac{k_{\min}(D)-1}{B_1-1},
\qquad
B(D)=\frac{U_2-k_{m_2}(D)}{U_2-L_2}.
\]
These correspond to the minimum-distance and second-moment terms in \(\Phi_\lambda(D)\), respectively.
For foldover designs, \(A(D)\) and \(B(D)\) have the desirability-scale
interpretation developed in Section~\ref{sec4}.
For nonfoldover designs, the same normalized quantities are used as common
benchmarked summaries.
Larger values of \(A(D)\) and \(B(D)\), and \(\Phi_\lambda(D)\) are preferred.

\subsection{Effect of the weight parameter}\label{subsec:lambda_sensitivity}

We first study how the weight parameter \(\lambda\) affects the two components of \(\Phi_\lambda(D)\). For \(m\in\{5,10,20\}\), run sizes \(n\in\{m,2m,3m,4m,5m\}\), and \(\lambda\in\{0,0.1,\ldots,0.9,1\}\), we construct 20 independent FSA-KD designs using 6000 simulated annealing iterations and 8 restarts.
For each design, we record the normalized minimum-distance component \(A(D)\) and the normalized second-moment component \(B(D)\).

\begin{figure}[ht!]
    \centering
    \includegraphics[width=0.98\textwidth]{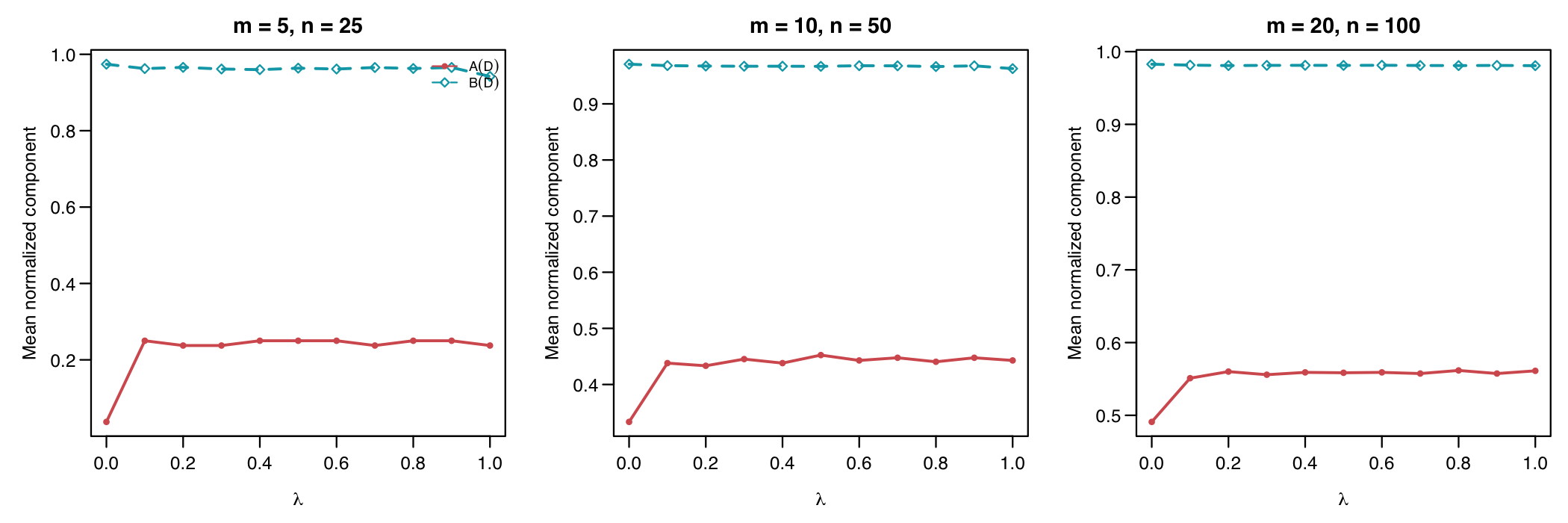}
    \caption{Effect of the weight parameter \(\lambda\) on the two normalized components of \(\Phi_\lambda(D)\) for \(n=5m\).  The red solid curves show the mean \(A(D)\), and the blue dashed curves show the mean \(B(D)\).  Points are averages over 20 independent FSA-KD runs for each \((m,\lambda)\).}
    \label{fig:wcrit_lambda_components}
\end{figure}

Figure~\ref{fig:wcrit_lambda_components} shows representative results for \(n=5m\), the densest run size considered in this experiment.
When \(\lambda=0\), the search is driven only by the second-moment component, and the resulting \(A(D)\) values are relatively small.
Once a positive weight is assigned to the minimum-distance component, \(A(D)\) increases markedly and then
changes only moderately over the remaining values of \(\lambda\).
The second-moment score \(B(D)\) remains close to one for all three values of \(m\), indicating that the improvement in worst-case Kendall tau distance does not come with a large loss in the moment component.
We therefore use \(\lambda=0.5\) as a balanced default in the remaining comparisons. The corresponding panels for \(n=m,2m,3m,\) and \(4m\) are provided in Section S1 of the Supplementary Materials.

\subsection{Comparison with maximin-\(L_2\) metaheuristics}\label{sec5-2}

\begin{figure}[ht!]
    \centering
    \includegraphics[width=0.9\textwidth]{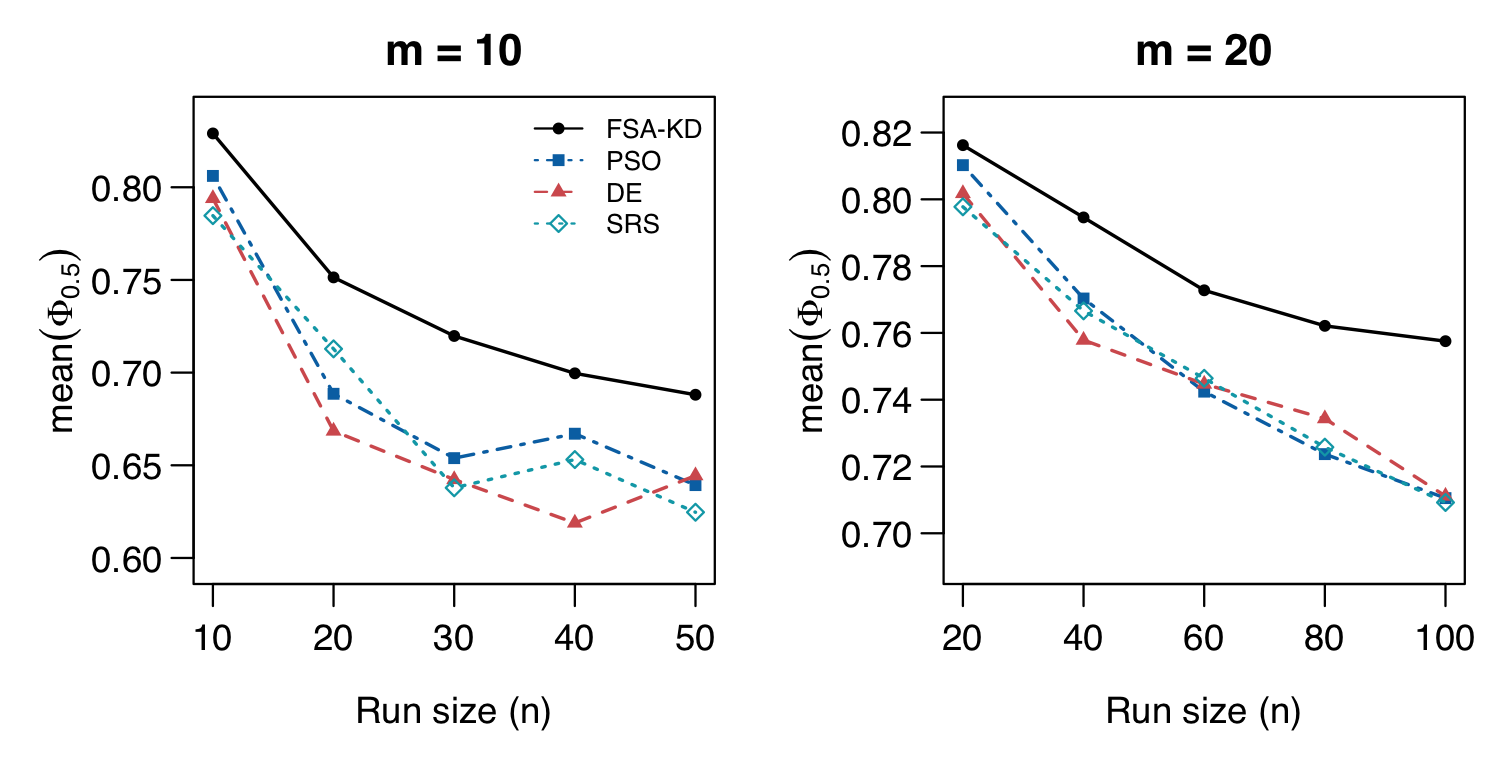}
    \caption{Comparison of FSA-KD, SRS, PSO, and DE under a common evaluation budget for \(m=10\) and \(m=20\). Each point is the mean \(\Phi_{0.5}(D)\) over 12 independent runs; larger values are preferred.}
    \label{fig:wcrit_algorithm_phi}
\end{figure}

We next compare FSA-KD with a representative maximin-\(L_2\) metaheuristic baseline 
from the \texttt{Meta4Design} R package, available in the online supplementary
materials of \citet{stokes2024metaheuristic}.
Following the comparison protocol in \citet{stokes2024metaheuristic}, the methods are run under a common evaluation budget.
We use \(m\in\{5,10,20\}\), \(n\in\{m,2m,3m,4m,5m\}\), and 12
independent repetitions for each setting. The target budget is 500 design criterion evaluations per repetition.
The DE and PSO implementations both optimize the same position-space maximin \(L_2\) criterion. They differ in the search heuristic but give similar Kendall tau criteria values in this simulation.
Table~\ref{tab:wcrit_multi_metric} therefore reports DE as a representative
maximin-\(L_2\) metaheuristic baseline, while
Figure~\ref{fig:wcrit_algorithm_phi} includes both DE and PSO.
All designs are then evaluated by the Kendall tau criteria and by \(\Phi_{0.5}(D)\).
For PSO and DE, we use population size \(NP=n\) and
\(\texttt{itermax}=\lfloor 500/n\rfloor\).
For FSA-KD, the same target budget is used as the total number of simulated-annealing iterations divided across the eight restarts. SRS has no iterative search budget.


Figure~\ref{fig:wcrit_algorithm_phi} compares  the methods using the weighted
Kendall tau criterion \(\Phi_{0.5}\).
FSA-KD gives the largest average value of \(\Phi_{0.5}\) in the plotted settings.
The curves decline as \(n\) increases, as expected, because close pairs are
harder to avoid when more runs are selected from the same finite permutation
space. A three-panel version is provided in Section S2 of the Supplementary
Material.

\begin{table}[ht!]
\centering
\caption{Comparison of DE-based maximin-\(L_2\) designs and FSA-KD designs under multiple design criteria for \(m=5\), \(m=10\), and \(m=20\).  Entries are averages over 12 independent runs.  Here \(d_E\) denotes the minimum Euclidean distance of the design. The average distance \(k_{\ave}\) is computed as in \eqref{kave}, and \(\Phi_{0.5}\) is used as a common evaluation score.  Bold entries indicate the larger values for \(d_E\), \(k_{\min}\), \(k_{\ave}\), and \(\Phi_{0.5}\), and the smaller value for \(k_{m_2}\), which are preferred.}
\label{tab:wcrit_multi_metric}
\scriptsize
\resizebox{\textwidth}{!}{%
\begin{tabular}{ccrrrrrrrrrr}
\toprule
 & & \multicolumn{5}{c}{DE} & \multicolumn{5}{c}{FSA-KD} \\
\cmidrule(lr){3-7} \cmidrule(lr){8-12}
\(m\) & \(n\) & \(d_E\) & \(k_{\min}\) & \(k_{\ave}\) & \(k_{m_2}\) & \(\Phi_{0.5}\) & \(d_E\) & \(k_{\min}\) & \(k_{\ave}\) & \(k_{m_2}\) & \(\Phi_{0.5}\) \\
\midrule
5 & 5 & \textbf{4.22} & 2.58 & 5.55 & \textbf{35.2} & 0.731 & 2.43 & \textbf{4.00} & \textbf{6.00} & 40.2 & \textbf{0.778} \\
5 & 10 & \textbf{2.80} & 1.17 & 5.16 & \textbf{30.4} & 0.586 & 1.85 & \textbf{4.00} & \textbf{5.56} & 33.9 & \textbf{0.827} \\
5 & 15 & \textbf{2.07} & 1.17 & 5.15 & \textbf{30.3} & 0.553 & 1.51 & \textbf{3.00} & \textbf{5.33} & 31.6 & \textbf{0.737} \\
5 & 20 & \textbf{1.51} & 1.00 & 5.09 & \textbf{29.9} & 0.529 & 1.46 & \textbf{2.00} & \textbf{5.26} & 31.1 & \textbf{0.610} \\
5 & 25 & \textbf{1.41} & 1.00 & 5.04 & \textbf{29.2} & 0.544 & \textbf{1.41} & \textbf{2.00} & \textbf{5.20} & 30.6 & \textbf{0.617} \\
10 & 10 & \textbf{10.07} & 10.50 & 22.38 & \textbf{531.0} & 0.794 & 7.96 & \textbf{18.42} & \textbf{25.00} & 680.9 & \textbf{0.829} \\
10 & 20 & \textbf{8.10} & 7.50 & 22.81 & \textbf{553.2} & 0.668 & 6.20 & \textbf{13.42} & \textbf{23.68} & 606.4 & \textbf{0.751} \\
10 & 30 & \textbf{7.01} & 5.92 & 22.41 & \textbf{533.3} & 0.642 & 5.51 & \textbf{11.50} & \textbf{23.28} & 583.2 & \textbf{0.720} \\
10 & 40 & \textbf{6.45} & 5.42 & 22.54 & \textbf{539.1} & 0.619 & 4.96 & \textbf{10.33} & \textbf{23.08} & 571.4 & \textbf{0.700} \\
10 & 50 & \textbf{6.05} & 6.58 & 22.53 & \textbf{538.3} & 0.644 & 4.43 & \textbf{9.67} & \textbf{22.96} & 564.6 & \textbf{0.688} \\
20 & 20 & \textbf{27.60} & 53.00 & 95.18 & \textbf{9293.5} & 0.802 & 24.31 & \textbf{69.33} & \textbf{100.00} & 10607.2 & \textbf{0.816} \\
20 & 40 & \textbf{24.35} & 47.42 & 95.23 & \textbf{9306.7} & 0.758 & 21.50 & \textbf{61.00} & \textbf{97.44} & 9933.8 & \textbf{0.795} \\
20 & 60 & \textbf{22.69} & 45.42 & 95.06 & \textbf{9271.9} & 0.745 & 20.78 & \textbf{55.33} & \textbf{96.61} & 9703.8 & \textbf{0.773} \\
20 & 80 & \textbf{21.95} & 43.83 & 95.02 & \textbf{9266.6} & 0.734 & 19.22 & \textbf{52.58} & \textbf{96.20} & 9593.4 & \textbf{0.762} \\
20 & 100 & \textbf{21.49} & 39.58 & 94.95 & \textbf{9253.4} & 0.711 & 18.53 & \textbf{51.25} & \textbf{95.96} & 9525.8 & \textbf{0.758} \\
\bottomrule
\end{tabular}%
}
\end{table}
Table~\ref{tab:wcrit_multi_metric} illustrates the difference between
optimizing a position-space maximin $L_2$-distance criterion (which is the minimum Euclidean distance of the design, denoted by $d_E$) and optimizing Kendall tau distance criteria.
As expected, DE usually has a larger \(d_E\), because it directly targets a maximin-\(L_2\) criterion.
By contrast, FSA-KD gives larger \(k_{\min}\) and \(k_{\ave}\) in every
reported setting, indicating better worst-case and average separation in pairwise precedence space.
DE often has a smaller \(k_{m_2}\), but this advantage does not offset its
weaker Kendall tau separation under the composite criterion for \(m=10\) and
\(m=20\).
For \(m=5\), the permutation space is small and the attainable distance
patterns are more limited; the composite scores are therefore closer.
Overall, the results show that a design optimized for position-space separation
need not be space-filling under the Kendall tau distance criteria, while FSA-KD targets
this geometry directly.

\subsection{Model-based validation}

The preceding comparisons concern design's space filling properties.  We now examine whether the same designs are useful in model-based tasks.  These experiments are motivated by the two statistical connections in Section~\ref{sec3}, the PWO model and the Mallows-kernel GP model.
The goal is not to claim uniform dominance in all prediction problems, but to assess whether the Kendall tau space filling can improve performance in representative downstream tasks.
Both prediction studies are reported using the normalized root mean squared prediction error (nRMSE), with smaller values preferred.
Specifically, for held-out true responses \(f_t\) and predictions \(\hat f_t\),
\[
\mathrm{nRMSE}
=
\frac{
\left\{N_{\mathrm{test}}^{-1}\sum_{t=1}^{N_{\mathrm{test}}}
(\hat f_t-f_t)^2\right\}^{1/2}
}{
\left\{(N_{\mathrm{test}}-1)^{-1}\sum_{t=1}^{N_{\mathrm{test}}}
(f_t-\bar f)^2\right\}^{1/2}
},
\]
where \(\bar f\) is the mean of the held-out true responses and $N_{\mathrm{test}}$ is the number of test set points.

We first consider prediction under a sparse PWO model.
We focus on the small-sample setting \(m=10\) and \(n/m\in\{2,3\}\), where the
number of PWO effects is \(q=\binom{10}{2}=45\) and regularization is needed to
stabilize the fit.
In each of 50 repetitions, the true PWO model has 10 active effects selected uniformly without replacement, with coefficients 
\(\beta_{ab}=s_{ab}u_{ab}\), where \(s_{ab}\in\{-1,1\}\) with equal probability
and \(u_{ab}\sim\mathrm{Unif}(1,2)\).  Gaussian noise \(\varepsilon \sim N(0,\, (\sigma_s/2)^2)\) is added, where \(\sigma_s\) is the standard deviation of the noiseless signal, giving a signal-to-noise ratio of 2.
The fitted model is an elastic net with mixing parameter \(\alpha=0.5\) using the \texttt{cv.glmnet} function in the \texttt{glmnet} R package, with an intercept, unstandardized PWO columns, five-fold cross-validation, mean squared error loss, and the conservative \(\lambda_{\mathrm{1se}}\) rule.
We compare FSA-KD designs with SRS designs and with maximin designs based on Hamming and
\(L_2\) distances.
The Hamming baseline is included as a standard distance criterion on orders,
whereas the \(L_2\) baseline links this validation to the position-distance
criterion used in Section~\ref{sec5-2}.
Prediction is evaluated on $2000$ held-out permutations (test set).

\begin{table}[ht!]
	\centering
	\caption{Average nRMSE for sparse PWO prediction with \(m=10\) under the conservative \(\lambda_{\mathrm{1se}}\) rule.  Entries are averages over 50 independent repetitions, with standard errors in parentheses.  Smaller values are better; the best value in each column is set in bold.}
	\label{tab:pwo_screening_summary}
	\begin{tabular}{lcc}
		\toprule
		Method & \(n=20\) & \(n=30\) \\
		\midrule
		FSA-KD                         & \textbf{0.835} (0.016) & \textbf{0.742} (0.017) \\
		Maximin-Hamming                & 0.838 (0.021) & 0.744 (0.021) \\
		Maximin-\(L_2\)                    & 0.859 (0.019) & 0.753 (0.019)\\
        SRS                            & 0.889 (0.018) & 0.745 (0.025) \\
		\bottomrule
	\end{tabular}
\end{table}

Table~\ref{tab:pwo_screening_summary} shows that FSA-KD designs have the smallest prediction error in both reported small-sample settings.
These results suggest that the proposed foldover Kendall tau construction can be useful when the PWO fit is regularized and the run size is limited.

We next evaluate the designs under a Mallows-kernel GP model.
We use a nonlinear job scheduling objective as a model-agnostic validation task. 
For an order \(\x=(x_1,\ldots,x_m)\), the objective function has the form \[
f(\x)=
\sum_{r=1}^m c_{x_r}\left(\sum_{\ell=1}^r p_{x_\ell}\right)^3
+\gamma s_J\{U(\x)-\bar U\}/s_U,
\]
where \(U(\x)=\sum_{a<b}\eta_{ab}\mathbf{1}\{a \hbox{ precedes } b \hbox{ in } \x\}\).
The processing times \(p_i\) and costs \(c_i\) are drawn once per repetition from truncated normal distributions, \(\eta_{ab}\) are independent standard normal coefficients, and \(s_J,\bar U,s_U\) are normalization constants computed from 5000 reference permutations.
We use \(\gamma=0.20\) for \(m=5\) and \(\gamma=0.30\) for \(m=10\).

Since the Mallows-kernel surrogate requires enough observations to
estimate the correlation parameter reliably, we report the settings
\(n=2m\) and \(n=4m\) for \(m\in\{5,10\}\).
These settings provide a useful test of whether the Kendall tau separation obtained by FSA-KD designs remains beneficial when the run size is small relative to the permutation space.
For each setting and 10 repetitions, we fit a
Mallows-kernel GP to each design, select the kernel parameter by marginal
likelihood over a fixed grid, and compute nRMSE on a held-out test set of 500 permutations.

\begin{table}[ht!]
\centering
\caption{Job-scheduling GP prediction results under the Mallows kernel.  Entries are mean nRMSE values over 10 independent repetitions, with standard errors in parentheses; smaller values are preferred.}
\label{tab:gp_fsakd}
\begin{tabular}{cccccc}
\toprule
\(m\) & \(n\) & FSA-KD & Maximin-Hamming & Maximin-\(L_2\) & SRS \\
\midrule
 5 & 10 & \textbf{0.375} (0.023) & 0.544 (0.034) & 0.507 (0.015) & 0.509 (0.027) \\
 5 & 20 & \textbf{0.242} (0.014) & 0.293 (0.018) & 0.307 (0.024) & 0.405 (0.018) \\
10 & 20 & \textbf{0.546} (0.032) & 0.554 (0.028) & 0.572 (0.015) & 0.609 (0.027) \\
10 & 40 & \textbf{0.423} (0.013) & 0.472 (0.014) & 0.493 (0.016) & 0.502 (0.017) \\
\bottomrule
\end{tabular}
\end{table}

For all four settings in Table~\ref{tab:gp_fsakd}, FSA-KD gives the lowest
mean nRMSE.
These results suggest that an initial design with good Kendall tau separation can improve the prediction capability of the Mallows-kernel surrogate, especially at relatively small run sizes.

Finally, we use the designs as initial samples in a Bayesian optimization
experiment for the same job scheduling objective.
Each run starts with \(n_{\rm init}=2m\) initial evaluations and then performs
expected-improvement (EI) steps under the Mallows-kernel GP surrogate as in \cite{deshwal2022}.
The GP is refit after each new observation.
The EI acquisition is optimized over unobserved permutations, with the search set consisting of all remaining permutations for \(m=5\) and a fresh random pool of 1000 unobserved permutations at each step for \(m=10\).
Thus repeat sampling is not allowed.
Because the objective is known in this validation study, the true optimum over
the full permutation space is computed for each repetition and shown after the
same scaling.


\begin{figure}[ht!]
    \centering
    \includegraphics[width=0.98\textwidth]{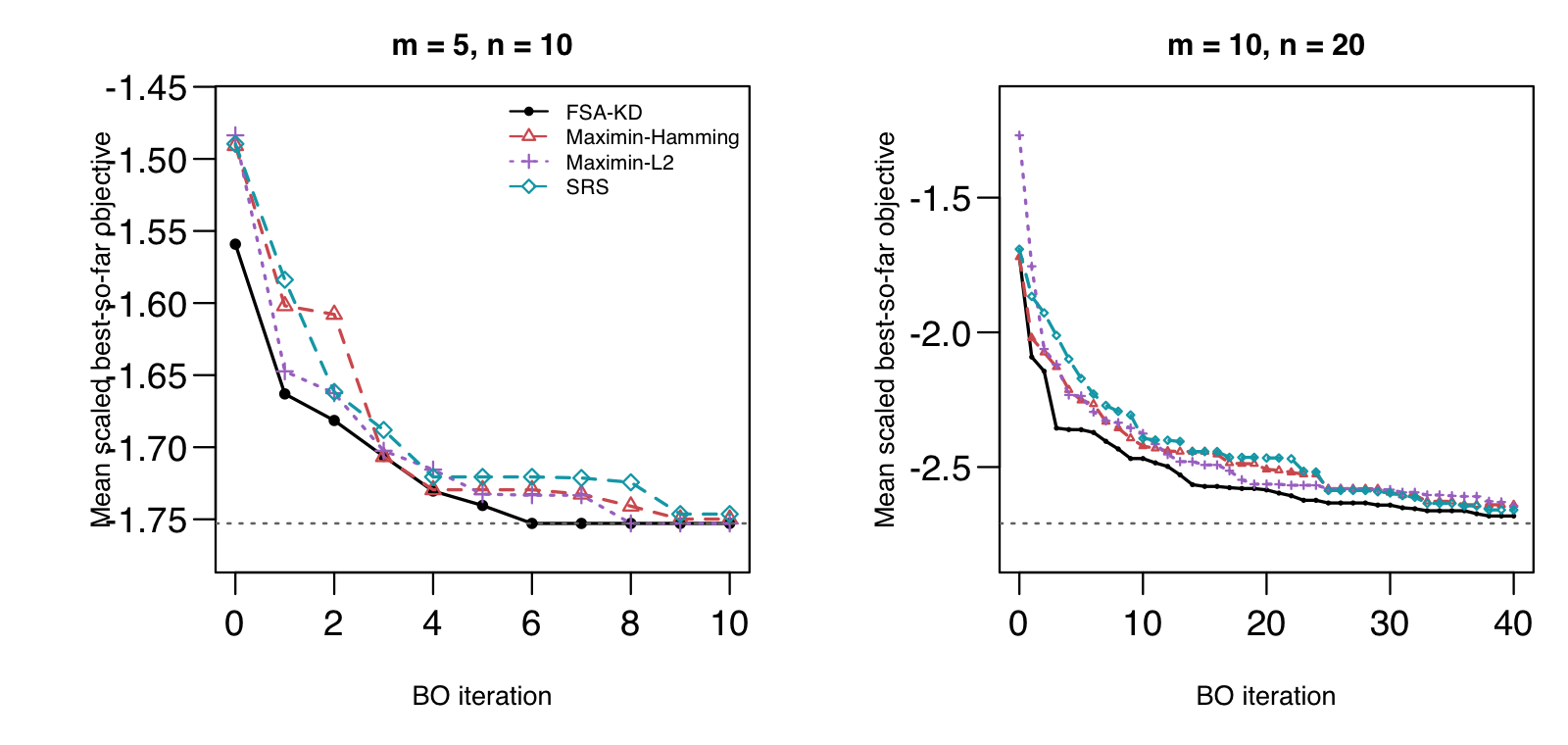}
    \caption{Bayesian optimization curves for the job scheduling objective with \(n_{\rm init}=2m\).  Left: \(m=5\), plotted through 10 BO iterations.  Right: \(m=10\), plotted through 40 BO iterations.  Curves are averages over 10 independent repetitions.  The vertical axis is the mean scaled best-so-far objective value; smaller values are better.  The gray horizontal dashed lines mark the mean scaled true optima over the full permutation space.}
    \label{fig:bo_validation}
\end{figure}

In the \(m=5\) panel of Figure~\ref{fig:bo_validation}, FSA-KD gives the best initial value, and its mean curve reaches the dashed optimum reference within the displayed iterations.
In the \(m=10\) panel, none of the methods reaches the true optimum by 40 iterations, but FSA-KD moves closest to the reference line after the first few expected-improvement steps.
The advantage of FSA-KD is most visible early in the search.
It provides a stronger starting design, after which the acquisition rule gradually reduces the differences among the initial designs.
Taken together, the results in this section suggest that the weighted criterion is useful for balancing worst-case separation motivated by the GP model and the dispersion of the pairwise distance distribution motivated by the PWO model.

\section{Concluding remarks}\label{sec6}

This paper studied space-filling OofA designs based on the Kendall tau distance.
Unlike position-based distances, the Kendall tau distance measures discrepancies in pairwise precedence relations and is therefore well aligned with the PWO representation of addition orders.
We considered the minimum distance, the average distance, and the second moment of the pairwise distance distribution as design criteria.
The connections with the MS criterion under the PWO model, the centralized generalized wordlength pattern, and the Mallows-kernel GP model provide statistical support for using these distance summaries in addition to their natural geometric interpretation.

We also propose FSA-KD, an efficient simulated annealing algorithm that uses a foldover structure and incremental distance updates for constructing space-filling OofA designs.
Theoretical results show that the search can be restricted to a representative half-design, 
while \(k_{\mathrm{ave}}\) is fixed and the completed distance distribution is
determined by that half-design. 
These properties motivate a well-defined weighted criterion that combines \(k_{\min}\) with \(k_{m_2}\) for controlling the dispersion of the pairwise distance distribution.
Numerical studies show that FSA-KD produces designs with favorable Kendall tau space-filling performance compared with random sampling and position-distance metaheuristic baselines of \cite{stokes2024metaheuristic}.
Our designs also perform well in the prediction and Bayesian optimization tasks over permutation spaces.

The proposed method is most useful when exact OofA-OAs or COAs are unavailable, or when the run size is determined by practical cost constraints. However, the FSA-KD algorithm may still face computational challenges because the permutation space still grows rapidly with the number of components, even after the foldover reduction.
More efficient implementations, parallel search strategies, and hybrid algorithms may further improve the construction for larger \(m\).
It would also be useful to develop algebraic constructions with good Kendall tau distance properties.
Other possible extensions include constrained orders, partial orders, and sequential designs that combine an initial FSA-KD design with adaptive sampling or Bayesian optimization \citep{stokes2024, XiaoWangMandalDeng24, rios2025}.

 \section{Disclosure statement}\label{disclosure-statement}

 The authors report there are no competing interests to declare.

 \section{Data availability statement}\label{data-availability-statement}

The data are included in the supplementary materials. 

\phantomsection\label{supplementary-material}
\bigskip
\begin{center}
{\large\bf SUPPLEMENTARY MATERIALS}
\end{center}

\begin{description}
\item[Supplementary materials:]
Additional sensitivity results, algorithm-comparison curves, a detailed
FSA-KD algorithm, and proofs of the theoretical results. (PDF)

\end{description}


\begin{thebibliography}{3}

        \bibitem[{Aarts and Korst(1988)}]{aarts1988}
        \textsc{Aarts, E.} and \textsc{Korst, J.} (1988).
        \newblock \textit{Simulated Annealing and Boltzmann Machines}.
        \newblock Chichester, UK: Wiley.


        \bibitem[{Deshwal et al.(2022)}]{deshwal2022}
        \textsc{Deshwal, A.}, \textsc{Belakaria, S.}, \textsc{Doppa, J. R.} and \textsc{Kim, D. H.} (2022).
        \newblock Bayesian optimization over permutation spaces.
        \newblock In \textit{Proceedings of the AAAI Conference on Artificial Intelligence}
        \textbf{36}(6), 6515--6523.

        \bibitem[{Eccleston and Hedayat(1974)}]{eccleston1974}
        \textsc{Eccleston, J. A.} and \textsc{Hedayat, A.} (1974).
        \newblock On the theory of connected designs: Characterization and optimality.
        \newblock \textit{The Annals of Statistics} \textbf{2}, 1238--1255.


        \bibitem[{Hedayat et al.(1999)}]{hedayat1999}
        \textsc{Hedayat, A. S.}, \textsc{Sloane, N. J. A.} and \textsc{Stufken, J.} (1999).
        \newblock \textit{Orthogonal Arrays: Theory and Applications}.
        \newblock New York, NY: Springer.


        \bibitem[{Huang and Yang(2025)}]{huang2025}
        \textsc{Huang, Y.} and \textsc{Yang, J.-F.} (2025).
        \newblock Robust design for order-of-addition experiments.
        \newblock \textit{Technometrics} \textbf{67}, 168--176.

        \bibitem[{Jiao and Vert(2015)}]{jiao2015}
        \textsc{Jiao, Y.} and \textsc{Vert, J.-P.} (2015).
        \newblock The Kendall and Mallows kernels for permutations.
        \newblock In \textit{Proceedings of the 32nd International Conference on Machine Learning},
        \textit{Proceedings of Machine Learning Research} \textbf{37}, 1935--1944.

        \bibitem[{Johnson et al.(1990)Johnson, Moore and Ylvisaker}]{johnson1990}
        \textsc{Johnson, M. E.}, \textsc{Moore, L. M.} and \textsc{Ylvisaker, D.} (1990).
        \newblock Minimax and maximin distance designs.
        \newblock \textit{Journal of Statistical Planning and Inference} \textbf{26}, 131--148.


        \bibitem[{Kendall(1938)}]{kendall1938}
        \textsc{Kendall, M. G.} (1938).
        \newblock A new measure of rank correlation.
        \newblock \textit{Biometrika} \textbf{30}, 81--93.

        \bibitem[{Kirkpatrick et al.(1983)Kirkpatrick, Gelatt and Vecchi}]{kirkpatrick1983}
        \textsc{Kirkpatrick, S.}, \textsc{Gelatt, C. D.} and \textsc{Vecchi, M. P.} (1983).
        \newblock Optimization by simulated annealing.
        \newblock \textit{Science} \textbf{220}, 671--680.

        \bibitem[{Li et al.(2022)}]{li2022}
        \textsc{Li, W.}, \textsc{Li, M.}, \textsc{Zhou, Y.} and \textsc{Yang, J.} (2022).
        \newblock Uniform order-of-addition designs (in Chinese).
        \newblock \textit{Scientia Sinica Mathematica} \textbf{52}, 1095--1112.

        \bibitem[{Lin and Peng(2019)}]{linpeng2019}
        \textsc{Lin, D. K. J.} and \textsc{Peng, J.} (2019).
        \newblock Order-of-addition experiments: A review and some new thoughts.
        \newblock \textit{Quality Engineering} \textbf{31}, 49--59.

        \bibitem[{Lin and Rios(2025)}]{linrios2025}
        \textsc{Lin, D. K. J.} and \textsc{Rios, N.} (2025).
        \newblock Order-of-addition experiments: A review and some recommendations.
        \newblock \textit{Wiley Interdisciplinary Reviews: Computational Statistics} \textbf{17}, e70024.


        \bibitem[{Mee(2020)}]{mee2020}
        \textsc{Mee, R. W.} (2020).
        \newblock Order of addition modeling.
        \newblock \textit{Statistica Sinica} \textbf{30}, 1543--1559.

        \bibitem[{Metropolis et al.(1953)Metropolis, Rosenbluth, Rosenbluth, Teller and Teller}]{metropolis1953}
        \textsc{Metropolis, N.}, \textsc{Rosenbluth, A. W.}, \textsc{Rosenbluth, M. N.}, \textsc{Teller, A. H.} and \textsc{Teller, E.} (1953).
        \newblock Equation of state calculations by fast computing machines.
        \newblock \textit{The Journal of Chemical Physics} \textbf{21}, 1087--1092.

        \bibitem[{Peng et al.(2019)Peng, Mukerjee and Lin}]{peng2019}
        \textsc{Peng, J.}, \textsc{Mukerjee, R.} and \textsc{Lin, D. K. J.} (2019).
        \newblock Design of order-of-addition experiments.
        \newblock \textit{Biometrika} \textbf{106}, 683--694.

        \bibitem[{Rios and Lin(2025)}]{rios2025}
        \textsc{Rios, N.} and \textsc{Lin, D. K. J.} (2025).
        \newblock Graphical methods for order-of-addition experiments.
        \newblock \textit{Journal of the Royal Statistical Society Series B: Statistical Methodology} \textbf{87}, 1309--1330.

        \bibitem[{Santner et al.(2018)Santner, Williams and Notz}]{santner2018}
        \textsc{Santner, T. J.}, \textsc{Williams, B. J.} and \textsc{Notz, W. I.} (2018).
        \newblock \textit{The Design and Analysis of Computer Experiments} (2nd ed.).
        \newblock New York, NY: Springer.


        \bibitem[{Schoen and Mee(2023)}]{schoenmee2023}
        \textsc{Schoen, E. D.} and \textsc{Mee, R. W.} (2023).
        \newblock Order-of-addition orthogonal arrays to study the effect of treatment ordering.
        \newblock \textit{The Annals of Statistics} \textbf{51}, 1877--1894.


        \bibitem[{Silvey(1980)}]{silvey1980}
        \textsc{Silvey, S. D.} (1980).
        \newblock \textit{Optimal Design}.
        \newblock London: Chapman and Hall.

        \bibitem[{Stokes and Xu(2022)}]{stokes2022}
        \textsc{Stokes, Z.} and \textsc{Xu, H.} (2022).
        \newblock A position-based approach for design and analysis of order-of-addition experiments.
        \newblock \textit{Statistica Sinica} \textbf{32}, 1467--1488.

        \bibitem[{Stokes and Xu(2024)}]{stokes2024}
        \textsc{Stokes, Z.} and \textsc{Xu, H.} (2024).
        \newblock Designs for order-of-addition screening experiments.
        \newblock \textit{Statistica Sinica} \textbf{34}, 399--419.

        \bibitem[{Stokes et al.(2024)Stokes, Wong and Xu}]{stokes2024metaheuristic}
        \textsc{Stokes, Z.}, \textsc{Wong, W. K.} and \textsc{Xu, H.} (2024).
        \newblock Metaheuristic solutions to order-of-addition design problems.
        \newblock \textit{Journal of Computational and Graphical Statistics} \textbf{33}, 1006--1016.

        \bibitem[{Tsai(2025)}]{tsai2025}
        \textsc{Tsai, S.-F.} (2025).
        \newblock Characterizing and comparing order-of-addition orthogonal arrays.
        \newblock \textit{Statistica Sinica} \textbf{37}(2), doi:10.5705/ss.202024.0191.

        \bibitem[{Van Nostrand(1995)}]{vannostrand1995}
        \textsc{Van Nostrand, R.} (1995).
        \newblock Design of experiments where the order of addition is important.
        \newblock In \textit{ASA Proceedings of the Section on Physical and Engineering Sciences}, pp. 155--160, Alexandria, VA: American Statistical Association.

        \bibitem[{Voelkel(2019)}]{voelkel2019}
        \textsc{Voelkel, J. G.} (2019).
        \newblock The design of order-of-addition experiments.
        \newblock \textit{Journal of Quality Technology} \textbf{51}, 230--241.


        \bibitem[{Wang and Lin(2023)}]{wanglin2023}
        \textsc{Wang, C.} and \textsc{Lin, D. K. J.} (2023).
        \newblock Interaction effects in pairwise ordering model.
        \newblock \textit{Journal of Quality Technology} \textbf{55}, 463--468.

        \bibitem[{Wang et al.(2020)Wang, Xu and Ding}]{wang2020}
        \textsc{Wang, A.}, \textsc{Xu, H.} and \textsc{Ding, X.} (2020).
        \newblock Simultaneous optimization of drug combination dose-ratio sequence with innovative design and active learning.
        \newblock \textit{Advanced Therapeutics} \textbf{3}, 1900135.

        \bibitem[{Wang et al.(2022)Wang, Xiao and Xu}]{wang2022}
        \textsc{Wang, Y.}, \textsc{Xiao, Q.} and \textsc{Xu, H.} (2022).
        \newblock On design orthogonality, maximin distance, and projection uniformity of computer experiments.
        \newblock \textit{Journal of the American Statistical Association} \textbf{117}, 375--385.


\bibitem[{Xiao and Xu(2021)}]{XiaoXu21}
\textsc{Xiao, Q.} and \textsc{Xu, H.} (2021).
\newblock A mapping-based universal Kriging model for order-of-addition experiments in drug combination studies.
\newblock \textit{Computational Statistics \& Data Analysis} \textbf{157}, 107155.

\bibitem[{Xiao et~al.(2024)}]{XiaoWangMandalDeng24}
\textsc{Xiao, Q.}, \textsc{Wang, Y.}, \textsc{Mandal, A.} and \textsc{Deng, X.} (2024).
\newblock Modeling and active learning for experiments with quantitative-sequence factors.
\newblock \textit{Journal of the American Statistical Association} \textbf{119}, 407--421.



\bibitem[{Xu and Wu(2001)}]{xuwu2001}
\textsc{Xu, H.} and \textsc{Wu, C.-F. J.} (2001).
\newblock Generalized minimum aberration for asymmetrical fractional factorial designs.
\newblock \textit{The Annals of Statistics} \textbf{29}, 1066--1077.

        \bibitem[{Yang et al.(2021)Yang, Sun and Xu}]{yang2021}
        \textsc{Yang, J.-F.}, \textsc{Sun, F.} and \textsc{Xu, H.} (2021).
        \newblock A component-position model, analysis and design for order-of-addition experiments.
        \newblock \textit{Technometrics} \textbf{63}, 212--224.


        \bibitem[{Zhao et al.(2021)Zhao, Lin and Liu}]{zhao2021}
        \textsc{Zhao, Y.}, \textsc{Lin, D. K. J.} and \textsc{Liu, M. Q.} (2021).
        \newblock Designs for order-of-addition experiments.
        \newblock \textit{Journal of Applied Statistics} \textbf{48}, 1475--1495.

\bibitem[{Zhou and Xu(2015)}]{ZhouXu15}
\textsc{Zhou, Y.} and \textsc{Xu, H.} (2015).
\newblock Space-filling properties of good lattice point sets.
\newblock \textit{Biometrika} \textbf{102}, 959--966.


      \end{thebibliography}
	

\end{document}


\begin{center}
{\LARGE\bf Supplementary Materials for}\\[0.5em]
{\LARGE \bf ``Space-filling foldover designs for order-of-addition experiments under Kendall tau distance criteria''}

\medskip

\medskip

{\large
Hui Shao, Yaping Wang, Dongdong Xiang, \\
KLATASDS--MOE, School of Statistics, \\
East China Normal University, Shanghai 200062, China \\
and \\
Qian Xiao \\
Department of Statistics, School of Mathematical Sciences, \\
Shanghai Jiao Tong University, Shanghai 200240, China \\
}

\end{center}

\bigskip
\spacingset{1.8}

\begin{description}
\item[Additional sensitivity results:]
Additional panels for the weight-parameter experiment in Section~5.1 of the main article.
\item[Algorithm-comparison curves:]
Curve-level comparison of FSA-KD, SRS, PSO, and DE under a common evaluation
budget.
\item[Detailed algorithm:]
An implementation-oriented version of the FSA-KD algorithm.
\item[Proofs:]
Proofs of the theoretical results in the main article.
\end{description}

\section{Additional Sensitivity Results}\label{app:sensitivity}

\begin{figure}[ht!]
    \centering
    \includegraphics[width=0.74\textwidth]{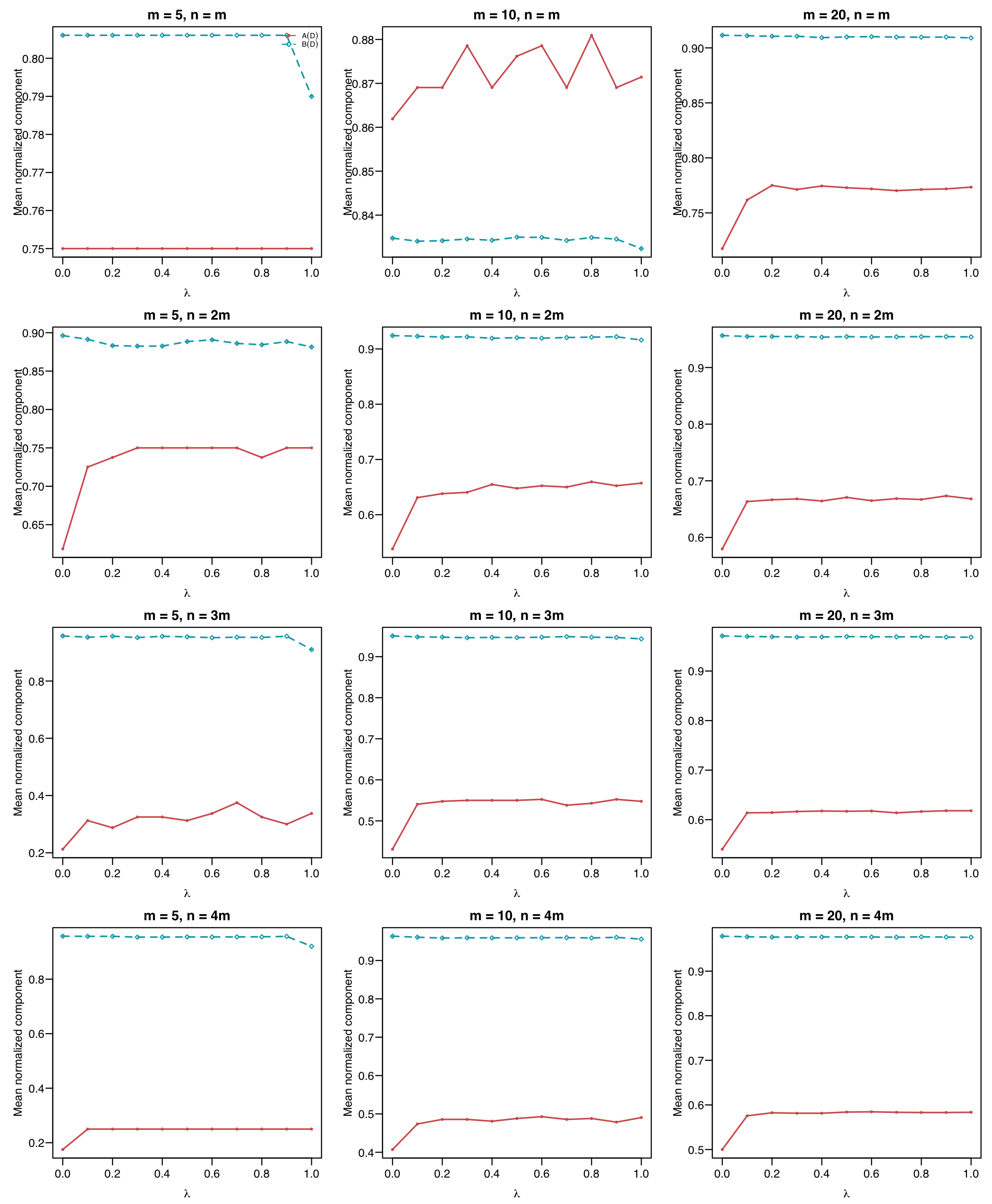}
    \caption{Supplementary sensitivity results for the weight parameter \(\lambda\) when \(n=m\), \(n=2m\), \(n=3m\), and \(n=4m\).  The red solid curve gives the normalized minimum-distance component \(A(D)\), and the blue dashed curve gives the normalized second-moment component \(B(D)\).}
    \label{fig:app_lambda_n1n4}
\end{figure}

Section~5.1 of the main article reports the sensitivity of the two normalized
components of \(\Phi_\lambda(D)\) for \(n=5m\).  The remaining run sizes from
the same experiment are shown in Figure~\ref{fig:app_lambda_n1n4}. The
experimental setup is identical to that in Figure~1 of the main article; only
the run-size multiplier \(n/m\) is changed.

The supplementary panels show the same qualitative trade-off as the \(n=5m\)
case in Figure~1 of the main article.  When \(\lambda=0\), the search is driven
entirely by the second-moment component and \(B(D)\) is close to one.  Assigning
positive weight to the minimum-distance component generally raises \(A(D)\),
while \(B(D)\) remains high across the reported settings.  The smallest designs
show less variation because the foldover structure already places strong
restrictions on the attainable distance distribution.

\section{Algorithm-Comparison Curves}\label{app:algorithm_phi}

\begin{figure}[ht!]
    \centering
    \includegraphics[width=0.95\textwidth]{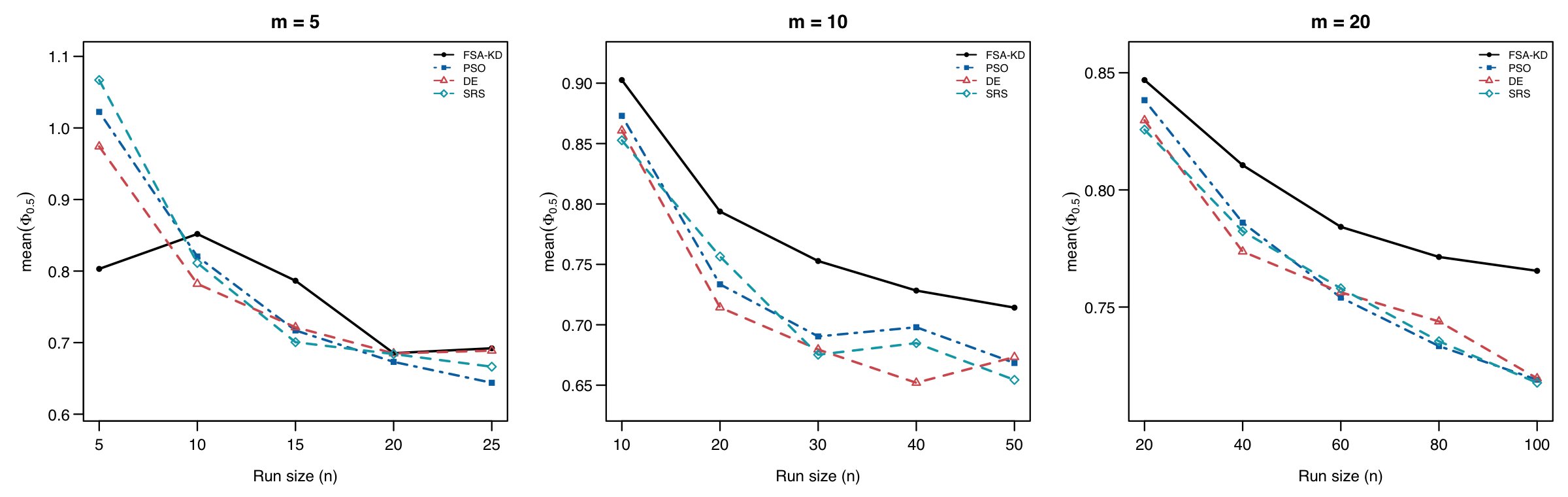}
    \caption{Supplementary comparison of FSA-KD, SRS, PSO, and DE under an equal-evaluations budget for \(m=5\), \(m=10\), and \(m=20\).}
    \label{fig:app_algorithm_phi}
\end{figure}

Section~5.2 of the main article reports the table-level comparison between
FSA-KD and a representative maximin-\(L_2\) metaheuristic baseline.  Figure
\ref{fig:app_algorithm_phi} gives the corresponding curve-level comparison of
FSA-KD, SRS, PSO, and DE under the same common-evaluation protocol.  We use
\(m\in\{5,10,20\}\), \(n\in\{m,2m,3m,4m,5m\}\), and 12 independent repetitions
for each setting.  The target budget is 500 design criterion evaluations per
repetition.  The PSO and DE implementations both optimize the same
position-space maximin \(L_2\) criterion, and all designs are evaluated by the
weighted Kendall tau criterion \(\Phi_{0.5}(D)\).

For \(m=10\) and \(m=20\), FSA-KD gives the largest average
\(\Phi_{0.5}\) across the displayed run sizes.  For \(m=5\), the methods are
closer because the permutation space is much smaller and fewer distance
patterns are attainable.  Across the panels, the curves tend to decline as
\(n\) increases, reflecting the fact that close pairs become harder to avoid
when more runs are placed in the same finite permutation space.

\section{Detailed FSA-KD Algorithm}

Algorithm~\ref{alg:SAKD-detail} gives an implementation-oriented version of
Algorithm~1 in the main article.  It makes explicit the two candidate-generation
moves and the incremental objective update used in our implementation.

\begin{algorithm}[ht!]
\caption{Detailed FSA-KD algorithm for space-filling foldover OofA designs}
\label{alg:SAKD-detail}
\begin{algorithmic}[1]
\Require Number of components $m$, even run size $n$ with $h=n/2$, initial temperature $T_0$, cooling factor $\alpha$, minimum temperature $T_{\min}$, maximum number of iterations $N_{\max}$, and weight parameter $\lambda$
\Ensure An optimized foldover OofA design $D^*$
\State Generate an initial representative half-design $H=\{\x_1,\ldots,\x_h\}$ such that $H\cup\widetilde H$ has no repeated run
\State Compute $\Phi_\lambda(D)$ for $D=H\cup\widetilde H$
\State Set $H^*\leftarrow H$ and $\Phi^*\leftarrow \Phi_\lambda(D)$
\State Set $T\leftarrow T_0$
\For{$t=1,\ldots,N_{\max}$}
    \If{$T<T_{\min}$}
        \State \textbf{break}
    \EndIf
    \State Randomly choose $r\in\{1,\ldots,h\}$
    \If{$\operatorname{rand}()<T/T_0$}
        \State Generate a new random permutation $\x'_r$
    \Else
        \State Generate $\x'_r$ by swapping two randomly chosen positions in $\x_r$
    \EndIf
    \State Form the candidate representative half-design $H'$
    \If{$H'\cup\widetilde H'$ has a repeated run}
        \State Reject $H'$
    \Else
        \State Compute $\Delta\Phi_\lambda
        =
        \Phi_\lambda(H'\cup\widetilde H')
        -
        \Phi_\lambda(H\cup\widetilde H)$ incrementally
        \If{$\Delta\Phi_\lambda>0$ \textbf{or} $\operatorname{rand}()<\exp\{\Delta\Phi_\lambda/T\}$}
            \State Set $H\leftarrow H'$
            \If{$\Phi_\lambda(H\cup\widetilde H)>\Phi^*$}
                \State Set $H^*\leftarrow H$ and $\Phi^*\leftarrow \Phi_\lambda(H\cup\widetilde H)$
            \EndIf
        \EndIf
    \EndIf
    \State Set $T\leftarrow \alpha T$
\EndFor
\State Construct $\widetilde H^*$ from $H^*$
\State Return $D^*=H^*\cup\widetilde H^*$
\end{algorithmic}
\smallskip \noindent\footnotesize\textit{Note.} Here $\operatorname{rand}()$ denotes an independent uniform draw from $[0,1]$.
\end{algorithm}

\clearpage

\section{Proofs}

\begin{proof}[Proof of Theorem 1]
By the definition of the MS-optimality function,
$$
\text{tr}(M^2) = \text{tr}(X^\top X X^\top X) = \text{tr}\left(  {X X^\top X X^\top} \right).
$$
We have
$$
X X^\top =
\left( \begin{array}{ccc}
1+z(\x_1)^\top z(\x_1) & \cdots & 1+z(\x_1)^\top z(\x_n) \\
\vdots & \ddots & \vdots \\
1+z(\x_n)^\top z(\x_1) & \cdots & 1+z(\x_n)^\top z(\x_n)
\end{array} \right).
$$
In the last equation, the diagonal elements are all equal to
\[
1+z(\x_i)^\top z(\x_i)  = \binom{m}{2} + 1, i=1,\ldots,n.
\]
For any two runs $\x_i$ and $\x_j$, each component pair contributes $+1$ to
$z(\x_i)^\top z(\x_j)$ when its relative order agrees in the two runs and $-1$
otherwise. Hence
\[
z(\x_i)^\top z(\x_j)= \binom{m}{2}-2k(\x_i,\x_j),
\]
and therefore
\[
k(\x_i,\x_j)=\frac12\left\{\binom{m}{2}-z(\x_i)^\top z(\x_j)\right\}.
\]
Then it follows that $\text{tr}(X X^\top X X^\top)$ is equal to
\Bea
& & 2\sum_{1\le i<j \le n} \left[ 1+z(\x_i)^\top z(\x_j) \right]^2 +\sum_{i=1}^n  \left[ 1+z(\x_i)^\top z(\x_i) \right]^2 \\
& = & 2\sum_{1\le i<j \le n} \left[ \binom{m}{2} + 1 - 2k(\x_i, \x_j) \right]^2 + n \left(\binom{m}{2}  +1\right)^2\\
& = & 2\sum_{1\le i<j \le n} \left[ 4k(\x_i, \x_j)^2 - 4\left(\binom{m}{2}  +1 \right) k(\x_i, \x_j) + \left(\binom{m}{2}  +1\right)^2 \right] + n\left(\binom{m}{2}  +1\right)^2\\
& = & 8\sum_{1\le i<j \le n} k(\x_i, \x_j)^2 - (4m(m-1)+8)\sum_{1\le i<j \le n} k(\x_i, \x_j) +  n^2 \left( \frac{m(m-1)}{2} + 1 \right)^2.
\Eea
By the definitions of $k_{\ave}(D)$ and $k_{m_2}(D)$ in the main article, the
conclusion follows.
\end{proof}

\begin{proof}[Proof of Proposition 1]
For $|W|=1$, the full design satisfies $J_W(D_{\mathrm{full}})=0$. Hence
\[
C_1(D)
=
\sum_{(a,b)\in\mathcal{Q}}
\left\{\frac{1}{n}\sum_{i=1}^n z_{ab}(\x_i)\right\}^2 =
\frac{1}{n^2}
\sum_{i=1}^n\sum_{j=1}^n
\sum_{(a,b)\in\mathcal{Q}}
z_{ab}(\x_i)z_{ab}(\x_j).
\]
For two runs $\x_i$ and $\x_j$,
\[
\sum_{(a,b)\in\mathcal{Q}}
z_{ab}(\x_i)z_{ab}(\x_j)
=
q-2k(\x_i,\x_j),
\]
because a component pair contributes $+1$ if its relative order agrees in the
two runs and $-1$ otherwise. Therefore
\[
C_1(D)
=
q-\frac{2}{n^2}
\sum_{i=1}^n\sum_{j=1}^n k(\x_i,\x_j)
=
q-\frac{2(n-1)}{n}k_{\ave}(D).
\]

For $|W|=2$, Theorem 3 of \citet{tsai2025} gives
\[
C_2(D)
=
\sum_{|W|=2}\left\{\frac{J_W(D)}{n}\right\}^2
-
\sum_{|W|=2}\left\{\frac{J_W(D_{\mathrm{full}})}{m!}\right\}^2 .
\]
For the full OofA design,
\[
\sum_{|W|=2}\left\{\frac{J_W(D_{\mathrm{full}})}{m!}\right\}^2
=
\frac{m(m-1)(m-2)}{18}.
\]
It remains to express the first term in distance form. Expanding the square,
\[
\sum_{|W|=2}\left\{\frac{J_W(D)}{n}\right\}^2
=
\frac{1}{n^2}
\sum_{i=1}^n\sum_{j=1}^n
\sum_{|W|=2}z_W(\x_i)z_W(\x_j).
\]
For a fixed ordered pair $(i,j)$, let $r=k(\x_i,\x_j)$. Among the $q$ pairwise
ordering factors, $q-r$ agree and $r$ differ between $\x_i$ and $\x_j$.
Therefore
\[
\sum_{|W|=2}z_W(\x_i)z_W(\x_j)
=
\binom{q-r}{2}+\binom{r}{2}-r(q-r)
=
\frac{q(q-1)}{2}-2qr+2r^2 .
\]
Summing over all ordered pairs gives
\[
\sum_{|W|=2}\left\{\frac{J_W(D)}{n}\right\}^2
=
\frac{q(q-1)}{2}
-\frac{2q(n-1)}{n}k_{\ave}(D)
+\frac{2(n-1)}{n}k_{m_2}(D),
\]
which completes the proof.
\end{proof}

\begin{proof}[Proof of Theorem 2]
For an OofA design $D$, write
$
\Sigma_D(\theta)=I_n+E_D(\theta),
$
where $E_D(\theta)$ has zero diagonal and off-diagonal entries
$
E_D(\theta)_{ij}
=
\exp\{-\theta k(\x_i,\x_j)\}, i\ne j.
$
Since the design has no replicated runs, $k_{\min}(D)>0$, and hence
$E_D(\theta)\to 0$ as $\theta\to\infty$.  Because $m$ and $n$ are fixed, the
set of candidate designs is finite, so the determinant expansion around the
identity is uniform over all candidate designs.  For a symmetric matrix $E$
with zero diagonal,
\[
\det(I_n+E)
=
1-\frac{1}{2}\operatorname{tr}(E^2)+O(\Vert E\Vert^3).
\]
Here $\Vert E\Vert$ denotes the spectral norm of $E$.
Applying this expansion to $E_D(\theta)$ gives
\begin{equation}\label{eq:det-expansion}
\det\{\Sigma_D(\theta)\}
=
1-
\sum_{1\le i<j\le n}
\exp\{-2\theta k(\x_i,\x_j)\}
+
O\{\exp[-3\theta k_{\min}(D)]\}.
\end{equation}
Equivalently,
\[
\det\{\Sigma_D(\theta)\}
=
1-
\sum_{r=k_{\min}(D)}^q
\nu_r(D)\exp(-2\theta r)
+
O\{\exp[-3\theta k_{\min}(D)]\}.
\]

If $k_{\min}(D)<k^*$, then the leading negative term in
\eqref{eq:det-expansion} for $D$ is of order
$\exp\{-2\theta k_{\min}(D)\}$, whereas the leading negative term for
$D^*$ is of smaller order $\exp(-2\theta k^*)$.  Therefore
\[
\det\{\Sigma_{D^*}(\theta)\}
>
\det\{\Sigma_D(\theta)\}
\]
for all sufficiently large $\theta$.

If $k_{\min}(D)=k^*$ but
$\nu_{k^*}(D)>\nu_{k^*}(D^*)$, then the two determinants have the same
exponential order in their leading terms, but the leading coefficient is
smaller for $D^*$.  Hence the same strict inequality holds for all
sufficiently large $\theta$. The desired result follows.
\end{proof}

\begin{proof}[Proof of Theorem 3]
(i).  Since
$z(\widetilde{\x})=-z(\x)$, reversing both runs leaves their PWO disagreements
unchanged.  Hence
\[
k(\widetilde{\x}_i,\widetilde{\x}_j)=k(\x_i,\x_j).
\]
Also, $\x_i$ and $\widetilde{\x}_i$ disagree on every component pair, so
$k(\x_i,\widetilde{\x}_i)=q$.  For $i\ne j$, a pairwise order agrees between
$\x_i$ and $\widetilde{\x}_j$ if and only if it disagrees between $\x_i$ and
$\x_j$.  Therefore
\[
k(\x_i,\widetilde{\x}_j)=q-k(\x_i,\x_j).
\]

(ii) Write
\[
S_1=\sum_{1\le i<j\le h}k(\x_i,\x_j),
\qquad
S_2=\sum_{1\le i<j\le h}k(\widetilde{\x}_i,\widetilde{\x}_j),
\]
and
\[
S_3=\sum_{i=1}^h\sum_{j=1}^h k(\x_i,\widetilde{\x}_j).
\]
By (i), $S_2=S_1$, and
\[
S_3
=
hq+\sum_{i\ne j}\{q-k(\x_i,\x_j)\}
=
h^2q-2S_1.
\]
Thus
\[
\sum_{1\le a<b\le n}k(\x_a,\x_b)
=
S_1+S_2+S_3
=
h^2q.
\]
Dividing by $\binom{2h}{2}$ and using $n=2h$ and $q=m(m-1)/2$ gives
\[
k_{\ave}(D)
=
\frac{h^2q}{\binom{2h}{2}}
=
\frac{nm(m-1)}{4(n-1)}.
\]

(iii) As $h\ge2$, part (i) gives
\[
k_{\min}(D)
=
\min_{1\le i<j\le h} \min
\{k(\x_i,\x_j),\,q-k(\x_i,\x_j)\}.
\]
For any integer $d$, $\min(d,q-d)\le \lfloor q/2\rfloor$.  Hence
\[
k_{\min}(D)\le \left\lfloor\frac{q}{2}\right\rfloor
=
\left\lfloor\frac{m(m-1)}{4}\right\rfloor.
\]
If $D$ has no repeated run, then all pairwise distances between distinct runs
are positive, and so $k_{\min}(D)\ge1$.

(iv) The lower bound follows from Corollary 1 and (ii). Substituting
\[
k_{\ave}(D)=\frac{nm(m-1)}{4(n-1)}
\]
into Corollary 1 gives
\[
k_{m_2}(D)\ge
\frac{nm(m-1)(9m^2-5m+10)}{144(n-1)}
=
L_2.
\]
Furthermore, Proposition 1 and the discussion in Section~3.2 of the main
article show that if $D$ is an OofA-OA$(n,m,2)$, then
$
k_{\ave}(D)
=
\frac{nm(m-1)}{4(n-1)},
$
and
$
k_{m_2}(D)
= L_2.
$

For the upper bound, let
\[
d_{ij}=k(\x_i,\x_j),\qquad 1\le i<j\le h.
\]
By part (i),
\[
\sum_{1\le a<b\le n}k^2(\x_a,\x_b)
=
hq^2+
2\sum_{1\le i<j\le h}\{d_{ij}^2+(q-d_{ij})^2\}.
\]
As $D$ has no repeated run, we have $1\le d_{ij}\le q-1$ for all $i<j$.  Hence
\[
d_{ij}^2+(q-d_{ij})^2
\le
1+(q-1)^2
=
q^2-2q+2.
\]
Therefore
\[
\sum_{1\le a<b\le n}k^2(\x_a,\x_b)
\le
hq^2+h(h-1)(q^2-2q+2).
\]
Dividing by $\binom{2h}{2}=h(2h-1)$ and substituting $h=n/2$ and
$q=m(m-1)/2$ gives
\[
k_{m_2}(D)
\le
\frac{n m^2(m-1)^2-4(n-2)\{m(m-1)-2\}}{8(n-1)}
=
U_2.
\]
This completes the proof.
\end{proof}